\documentclass[nofootinbib,reprint,floatfix, longbibliography, superscriptaddress]{revtex4-1}
\usepackage{amsmath}
\usepackage{amssymb}
\usepackage{graphicx}
\usepackage{verbatim}
\usepackage{color}
\usepackage{subfigure}
\usepackage{appendix}
\usepackage[pdftex, breaklinks,colorlinks, citecolor=blue, urlcolor=blue]{hyperref}

\usepackage{color}
\usepackage{subfigure}

\newcommand{\appref}[1]{\hyperref[#1]{{Appendix~\ref*{#1}}}}
\newcommand{\be}{\begin{eqnarray} \begin{aligned}}
\newcommand{\ee}{\end{aligned} \end{eqnarray} }
\newcommand{\benn}{\begin{eqnarray*} \begin{aligned}}
\newcommand{\eenn}{\end{aligned} \end{eqnarray*}}

\newcommand*{\cN}{\mathcal{N}}
\newcommand*{\cD}{\mathcal{D}}

\newcommand{\bc}{\begin{center}}
\newcommand{\ec}{\end{center}}

\newcommand{\ket}[1]{|#1\rangle}
\newcommand{\bra}[1]{\langle#1|}
\def\>{\rangle}
\def\<{\langle}

\newcommand{\proj}[1]{|#1\rangle\!\langle#1|}

\newcommand{\braket}[2]{\langle #1|#2\rangle}
\newcommand{\specificthanks}[1]{*}

\begin{document}

\title{Harnessing electro-optic correlations in an efficient mechanical converter}
\author{A. P. Higginbotham}
\thanks{These authors contributed equally}
\email[direct correspondence to ]{Peter.S.Burns@colorado.edu}
\author{P. S. Burns}
\thanks{These authors contributed equally}
\email[direct correspondence to ]{Peter.S.Burns@colorado.edu}
\author{M. D. Urmey}
\thanks{These authors contributed equally}
\email[direct correspondence to ]{Peter.S.Burns@colorado.edu}
\author{R. W. Peterson}
\author{N. S. Kampel}
\author{B. M. Brubaker}
\affiliation{JILA, National Institute of Standards and Technology and University of Colorado, and Department of Physics, University of Colorado, Boulder, Colorado 80309, USA}
\author{G. Smith}
\affiliation{JILA, National Institute of Standards and Technology and University of Colorado, and Department of Physics, University of Colorado, Boulder, Colorado 80309, USA}
\affiliation{Center for Theory of Quantum Matter, University of Colorado, Boulder, Colorado 80309, USA}

\author{K. W. Lehnert}
\author{C. A. Regal}
\affiliation{JILA, National Institute of Standards and Technology and University of Colorado, and Department of Physics, University of Colorado, Boulder, Colorado 80309, USA}

\begin{abstract} An optical network of superconducting quantum bits (qubits) is an appealing platform for quantum communication and distributed quantum computing, but developing a quantum-compatible link between the microwave and optical domains remains an outstanding challenge. Operating at $T < 100$~mK temperatures, as required for quantum electrical circuits, we demonstrate a mechanically-mediated microwave-optical converter with 47$\%$ conversion efficiency, and use a classical feedforward protocol to reduce added noise to 38~photons. The feedforward protocol harnesses our discovery that noise emitted from the two converter output ports is strongly correlated because both outputs record thermal motion of the same mechanical mode. We also discuss a quantum feedforward protocol that, given high system efficiencies, would allow quantum information to be transferred even when thermal phonons enter the mechanical element faster than the electro-optic conversion rate.  
\end{abstract}

\maketitle

A quantum network entangling spatially separated quantum nodes would permit secure communication and distributed quantum computing \cite{cirac_quantum_1997,ekert_quantum_1991,cirac_distributed_1999,Dur_entanglement_2003}. 
Propagating optical fields are the natural choice for quantum links in such a network, as they enable entanglement distribution at room temperature and over kilometer-scale distances.
The fast-paced development of superconducting quantum processors \cite{kelly_state_2015,ofek_extending_2016} suggests that the most powerful nodes will use microwave-frequency excitations in ultralow-temperature  environments ($T<100~\text{mK}$), implying that a quantum state preserving electro-optic converter is a crucial element needed for a future quantum internet.
Prospective conversion technologies under investigation are ultracold atoms~\cite{verdu_strong_2009,hafezi_atomic_2012}, optically active spins in solids~\cite{imamoglu_cavity_2009,marcos_coupling_2010,kubo_strong_2010,williamson_magneto-optic_2014}, magnons \cite{hisatomi_bidirectional_2016}, electro-optic materials~\cite{ilchenko_whispering-gallery-mode_2003,
xiong_low-loss_2012,rueda_efficient_2016}, and mechanical resonators~\cite{bochmann_nanomechanical_2013,andrews_bidirectional_2014,vainsencher_bi-directional_2016}. To date there has been no successful demonstration of an electro-optic converter capable of quantum operation. Recent realizations have demonstrated improved capacity for classical signal recovery, albeit at elevated operating temperatures\cite{andrews_bidirectional_2014,bochmann_nanomechanical_2013,bagci_optical_2014,vainsencher_bi-directional_2016,takeda_electro-mechano-optical_2018}.  

Realizing a quantum electro-optic converter is a challenging task because it entails bringing together superconducting quantum circuits and laser light in an ultralow-temperature environment. 
The potential for quantum operation can be delineated via two metrics: the bidirectional efficiency between the microwave and optical ports, and the added noise of the converter $N_{\mathrm{add}}$ \cite{safavi-naeini_proposal_2011,hill_coherent_2012,andrews_bidirectional_2014,zeuthen_figures_2016}. 
In a mechanical converter, low added noise requires high electromechanical and optomechanical cooperativities~\cite{safavi-naeini_design_2010,hill_coherent_2012}, which have yet to be achieved together in a single device. Alternatives to high cooperativity have been explored in other optomechanical systems; in particular, feedback damping has been experimentally shown to ease cooperativity requirements in individual platforms
\cite{cohadon_cooling_1999,arcizet_high-sensitivity_2006,poggio_feedback_2007,genes_ground-state_2008,wilson_measurement-based_2015,rossi_enhancing_2017,rossi_measurement-based_2018}.
But quantum electro-optic conversion studies have to date focused on reaching the threshold $N_{\mathrm{add}} < 1$, at which point arbitrarily low efficiency can be tolerated using quantum repeater concepts that herald the creation of entanglement probabilistically~\cite{duan_long_2001}.

In this work, we explore electro-optic conversion in an alternative feedforward framework where quantum tasks could be performed even if thermal-mechanical noise yields $N_{\mathrm{add}} > 1$, provided threshold efficiencies are reached. 
We demonstrate an unprecedented conversion efficiency of $47 \pm 1\%$ in a micromechanical device operated at $T<100~\mathrm{mK}$, and implement a feedforward protocol that exploits noise correlations between the two converter output ports to reduce added noise to $N_\mathrm{add} = 38~\mathrm{photons}$. These figures of merit represent significant technical progress relative to the $8\%$ efficiency and 1500~photons of added noise achieved in a prototype system operated at 4~K~\cite{andrews_bidirectional_2014}. Furthermore, while we focus on electro-optic converters here, our feedforward protocol is broadly applicable to two-mode signal processing devices in which transmission occurs through a noisy intermediary mode. 
It also complements recent theoretical work proposing adaptive control for quantum transducers~\cite{rakhubovsky_squeezer_2016,zhang_quantum_2018}.

\begin{figure}[h]
\begin{minipage}{\linewidth}
\scalebox{1}{\includegraphics{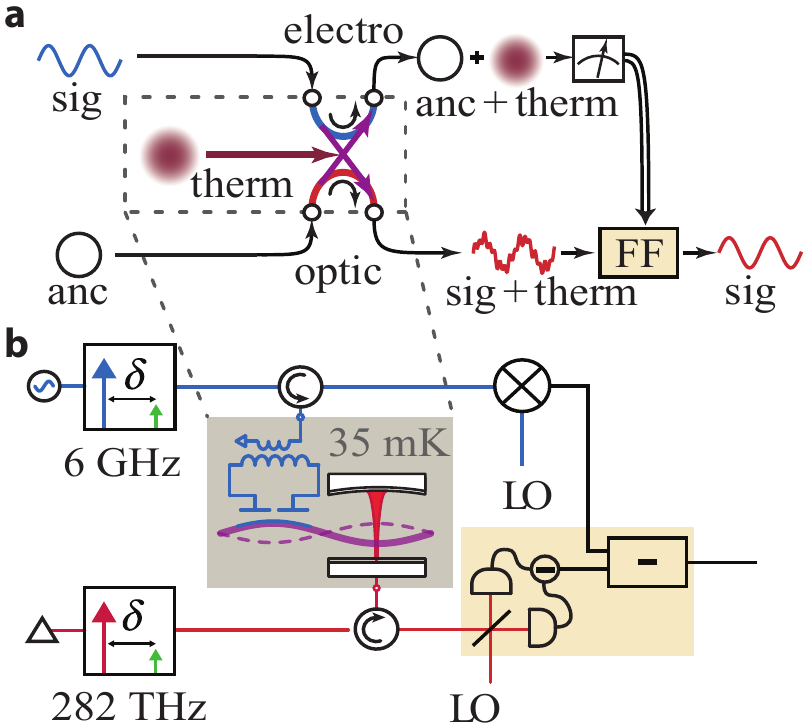}}
\caption{\textbf{Feedforward schematic and measurement network. a,} A signal incident on the microwave port (blue sinusoid) is output from the optical port (red sinusoid) with thermal noise coupled in from the internal port (maroon disk).
Optical ancilla (black circle) is simultaneously downconverted with correlated noise added. The ancilla is measured and a feedforward protocol (FF) is applied.
Dotted line is the converter box. Blue and red arrows show microwave and optical reflections, and purple arrows show conversion process.
\textbf{b,} Simultaneous coupling of microwave (blue) and optical (red) resonators to a single mechanical mode (purple).
Strong optical (red arrow) or microwave (blue arrow) pumps are applied to create optomechanical interaction.
A weak probe signal (green arrow) detuned from the pump by $\delta$ can be applied to one converter port and detected at the other.
Converter box is shaded grey. Feedforward operations are indicated by beige-shaded regions.
For our classical demonstration these operations are optical heterodyne measurement and subtraction in post-processing. The corresponding operation in a quantum feedforward protocol would be a unitary displacement of the optical field conditioned on the microwave measurement result.
}
\label{schematic1}
\end{minipage}
\end{figure}

We will consider classical and quantum feedforward protocols in a microwave-mechanical-optical converter in which the mechanical element is thermally occupied (Fig.~\ref{schematic1}a).  
The noise emitted from each port contains a redundant record of the mechanical oscillator's thermally driven motion. 
During operation, the signal to be converted is injected into the microwave port and an ancilla state is injected into the optical port.
The upconverted signal, with noise added, is emitted from the optical port.
The ancilla state, contaminated with the same added noise, is emitted from the microwave port.
Measuring the downconverted ancilla and feeding forward to the propagating optical mode can remove this correlated added noise.

Classical and quantum feedforward protocols are distinguished by the choice of ancilla.
Simply choosing the ancilla to be optical vacuum causes zero-point fluctuations (vacuum noise) to be fed forward along with the thermal noise.
In this case, although thermal noise may be completely removed, vacuum noise from the ancilla is necessarily written onto the upconverted signal.
Choosing vacuum as the ancilla therefore prohibits upconversion of a state with negative Wigner function or squeezing, so we refer to it as classical feedforward.
Classical feedforward is a resource for recovering classical signals, whose performance --- interestingly --- depends on system efficiencies rather than temperature, quality factor, or cooperativity.
However, classical feedforward is apparently unhelpful for quantum tasks.

A different choice of ancilla would allow quantum tasks to be accomplished, even in the presence of thermal noise.
For example if the ancilla is an infinitely squeezed vacuum state, one quadrature can be fed forward noiselessly, permitting noiseless measurement of a single upconverted quadrature or upconversion of a squeezed quadrature without high cooperativity.
This quantum protocol does however place limits on measurement and converter efficiencies, as any loss will add noise to the feedforward process.
In the limit of perfect measurement efficiencies and squeezing, a converter efficiency $\eta > 50\%$ is still required to upconvert a squeezed quadrature or measure a remotely prepared microwave quadrature with added noise less than vacuum.
We therefore refer to $50\%$ as a quantum threshold efficiency for feedforward protocols.
More sophisticated tasks can be imagined with quantum feedforward, albeit with more stringent efficiency requirements. In particular, quantum feedforward could be used to upconvert a qubit state given a qubit encoding robust to noise in one quadrature. One example of such a feedforward-based qubit upconversion scheme is explored in the supplement.


In the present work we focus on the experimental identification of correlations that allow feedforward, and perform classical feedforward with a vacuum ancilla.
For this demonstration we measure both converter outputs in heterodyne detection and perform subtraction in post-processing to remove the correlated noise (Fig.~\ref{schematic1}b). A quantum protocol would instead entail homodyne measurement of the ancilla at the microwave port, and conditional unitary displacement at the optical port.\cite{braunstein_teleportation_1998}

Our electro-optic converter is housed in a cryostat with base temperature $T=35~\mathrm{mK}$, and comprises microwave and optical cavities simultaneously coupled to a single vibrational mode of a suspended dielectric membrane with resonant frequency $\omega_\mathrm{m} / 2 \pi = f_\mathrm{m} = 1.473~\mathrm{MHz}$ (Fig.~\ref{schematic1}b).
Strong red-detuned pump tones incident on both cavities parametrically couple mechanical motion to propagating microwave and optical fields at rates $\Gamma_\mathrm{e,o}$ that greatly exceed the intrinsic mechanical damping rate $\gamma_\mathrm{m} = 2\pi\times11~\mathrm{Hz}$ (see methods).
Along with the two pump tones, a weak probe tone, incident on either the microwave or optical cavity, is measured in heterodyne detection, and used for characterization and implementation of classical feedforward. 

\begin{figure}
\begin{minipage}{\linewidth}
\scalebox{1}{\includegraphics{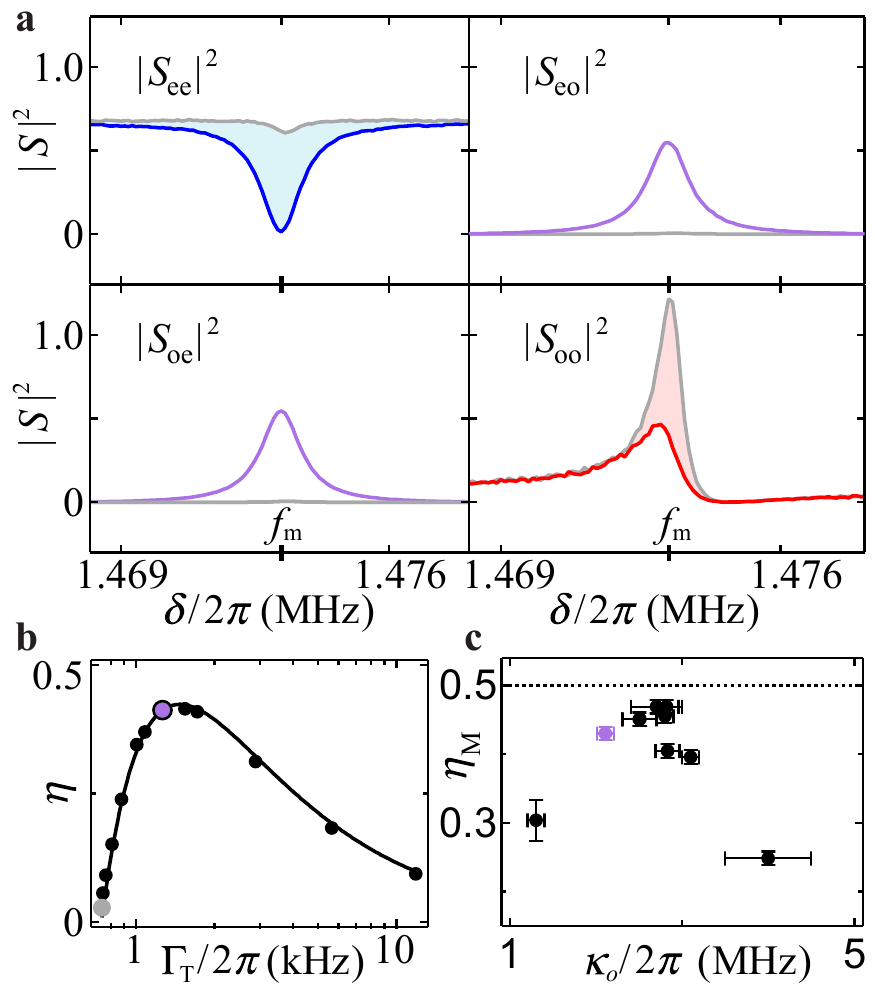}}
\caption{\textbf{Converter efficiency. a,} Measured converter scattering parameters versus probe frequency $\delta$: microwave reflection ($S_\mathrm{ee}$), optical reflection ($S_\mathrm{oo}$), microwave-to-optical transmission ($S_\mathrm{oe}$), and optical-to-microwave transmission ($S_\mathrm{eo}$). 
Colored trace is $\Gamma_\mathrm{e} \approx \Gamma_\mathrm{o}$ (see methods).
Gray trace is $\Gamma_\mathrm{e} \ll \Gamma_\mathrm{o}$, with same $\Gamma_\mathrm{o}$.
Shaded region highlights induced absorption of incident power when conversion rates are matched.
\textbf{b,} Converter efficiency $\eta$ versus total damping  $\Gamma_\mathrm{T}$. 
$\Gamma_\mathrm{T}$ is swept by tuning $\Gamma_\mathrm{e}$ with $\Gamma_\mathrm{o}$ fixed. Black line is a fit to Eq.~\ref{eq:etam}.
Purple and gray points correspond to data in panel \textbf{a}. 
\textbf{c,} Matched efficiency $\eta_\mathrm{M}$ versus optical cavity linewidth $\kappa_\mathrm{o}$.
The ratio of external optical cavity coupling to internal loss varies with $\kappa_\mathrm{o}$.
Black dotted line indicates quantum feedforward threshold at $\eta_\mathrm{M} = 0.5$, discussed in main text.
Horizontal error bars represent the standard deviation of several repeated linewidth measurements.
Vertical error bars are obtained by propagating standard error in individual scattering parameter measurements.
}
\label{sparam}
\end{minipage}
\end{figure}

The converter box is characterized by a series of probe-tone scattering parameter measurements $S_{ij}(\delta)$, where $i=\mathrm{e},\mathrm{o}$ is the measured port, $j=\mathrm{e},\mathrm{o}$ is the excited port, and $\delta$ is the frequency of the probe relative to the pump~\cite{andrews_bidirectional_2014}.
As shown in Fig.~\ref{sparam}a, when $\Gamma_\mathrm{e} \approx \Gamma_\mathrm{o}$, a dip in microwave reflection occurs near $\delta/2\pi=f_\mathrm{m}$, with a corresponding peak in microwave-to-optical transmission, indicating the absorption of signals in the microwave port and their emission at the optical port.
A nearly identical optical-to-microwave transmission signal is also observed.
The peak transmission $|t|^2 = 0.55$ corresponds to a conversion efficiency of $\eta = |t|^2/\mathcal{A} = 0.41$, where $\mathcal{A}$ is the independently measured converter gain due to imperfect sideband resolution~\cite{andrews_bidirectional_2014}. The converter bandwidth (measured as the full width at half maximum of the transmission peak) is given by the total linewidth of the optomechanically and electromechanically damped membrane mode, $\Gamma_\mathrm{T} = \Gamma_\mathrm{e} + \Gamma_\mathrm{o} + \gamma_\mathrm{m}$.

For comparison, if $\Gamma_\mathrm{e}$ is decreased such that $\Gamma_\mathrm{e} \ll \Gamma_\mathrm{o}$, microwave reflection becomes almost flat, with a value determined by the microwave cavity coupling, and nearly zero power is transmitted.
At the same time, a peak is observed in optical reflection resulting from optomechanically induced transparency effects. 
The peak height exceeds one due to converter gain (see supplement).
The optical reflection peak is suppressed when $\Gamma_\mathrm{e} \approx \Gamma_\mathrm{o}$, constituting electromechanically induced optical absorption, which, to our knowledge, has not been previously reported.

To further explore the converter box's performance, the efficiency is extracted from peak transmission for a range of $\Gamma_\mathrm{e}$ with $\Gamma_\mathrm{o} = 2 \pi \times 725~\mathrm{Hz}$ fixed (Fig.~\ref{sparam}b).
For fixed optical cavity parameters, efficiency is maximized when damping rates are matched ($\Gamma_\mathrm{e} = \Gamma_\mathrm{o}$, $\Gamma_\mathrm{T} = 2\pi \times 1.45 ~\mathrm{kHz}$).
The conversion efficiency is fit to \cite{andrews_bidirectional_2014}
\begin{equation} \label{eq:etam}
\eta = \frac{ 4 \Gamma_\mathrm{e} \Gamma_\mathrm{o} }{ (\Gamma_\mathrm{e} + \Gamma_\mathrm{o} + \gamma_\mathrm{m})^2 } \eta_\mathrm{M} = 
\frac{ 4 (\Gamma_\mathrm{T} - \Gamma_\mathrm{o} - \gamma_\mathrm{m})  \Gamma_\mathrm{o} }{ \Gamma_\mathrm{T}^2 } \eta_\mathrm{M},
\end{equation}
where $\gamma_\mathrm{m}$ and $\Gamma_\mathrm{o}$ are fixed from independent measurements and $\eta_\mathrm{M}$ is the only fit parameter. 
For our converter $\Gamma_\mathrm{e},\Gamma_\mathrm{o} \gg \gamma_\mathrm{m}$, and thus  $\eta=\eta_\mathrm{M}$ when the converter is matched. Therefore we refer to $\eta_M$ as the matched efficiency.
In this low mechanical dissipation regime, there is negligible energy loss in the electro-optic transduction process itself, but some energy is absorbed in each electromagnetic resonator, and the spatial profile of the optical cavity mode does not perfectly match that of the external modes used for measurement and signal injection (see supplement). 
We thus expect $\eta_\mathrm{M} = \epsilon (\kappa_\mathrm{ex,o}/\kappa_\mathrm{o})( \kappa_\mathrm{ex,e}/\kappa_\mathrm{e})$, 
where $\kappa_\mathrm{o}$ and $\kappa_\mathrm{ex,o}$ ($\kappa_\mathrm{e}$ and $\kappa_\mathrm{ex,e}$) are the optical (microwave) cavity linewidth and external coupling respectively, 
and $\epsilon$ parameterizes the optical cavity mode matching. 
The fit result, $\eta_\mathrm{M} = 43 \pm 1\%$, agrees with the theoretical expectation, $\eta_\mathrm{M} = 43 \pm 4\%$ obtained from independent measurements of $\epsilon$, $\kappa_\mathrm{ex,o}/\kappa_\mathrm{o}$, and $\kappa_\mathrm{ex,e}/\kappa_\mathrm{e}$.

Tuning the membrane position \textit{in situ} within the optical cavity changes both  $\kappa_\mathrm{ex,o}$ and $\kappa_\mathrm{o}$ due to interference effects (see supplement), and therefore alters the matched conversion efficiency (Fig.~\ref{sparam}c). 
Matched conversion efficiency initially increases with $\kappa_\mathrm{o}$, reaching a maximum at $\kappa_\mathrm{o}=2 \pi \times 2.7~\mathrm{MHz}$, and then decreases as internal optical cavity loss begins to dominate.
The \textit{in situ} tuning of the optical cavity also changes the optomechanical coupling and thus the achievable converter bandwidth $\Gamma_\mathrm{T}$.
For intermediate linewidths near the highest efficiencies, the optical cavity became unstable, possibly due to large optomechanical coupling in these regions. 
The peak conversion efficiency achieved was $47\pm1\%$, 
approaching the quantum feedforward threshold efficiency.
At peak conversion efficiency, the matched converter bandwidth was $12~\mathrm{kHz}$, while a $100~\mathrm{kHz}$ bandwidth was achieved at $\kappa_\mathrm{o} = 2 \pi \times 3~\mathrm{MHz}$.

During the conversion process, vibrational noise is added to the signal.
In order to explore correlations in this noise, we turn off the weak probe tone and return to a high-stability configuration with  $\eta_\mathrm{M} = 43\%$, but smaller pump power ($\Gamma_\mathrm{e} \approx \Gamma_\mathrm{o}$, $\Gamma_\mathrm{T} = 2 \pi \times 200~\mathrm{Hz}$). 
In this configuration electro/optomechanical cooling of the mechanical mode is diminished, while still maintaining $\Gamma_\mathrm{e},\Gamma_\mathrm{o} \gg \gamma_\mathrm{m}$. 
We perform heterodyne measurements of the noise exiting both converter ports (see supplement), and analyze the results in the frequency domain: $X(\omega)$ and $Y(\omega)$ will denote respectively the real and imaginary parts of the Fourier transform of the noise time stream at detuning $+\omega$ from the pump. The total power spectral density is then $S(\omega) = \langle X^2(\omega) \rangle + \langle Y^2(\omega) \rangle$, and is reported in units of $\mathrm{photons/s/Hz}$, or more simply photons, referred to the converter output.
For example, an ideal heterodyne measurement of $S( \omega )$ gives a background noise level of 1 photon, and $50\%$ loss in the measurement chain would double the background level.

\begin{figure}
\begin{minipage}{\linewidth}
\scalebox{.9}{\includegraphics{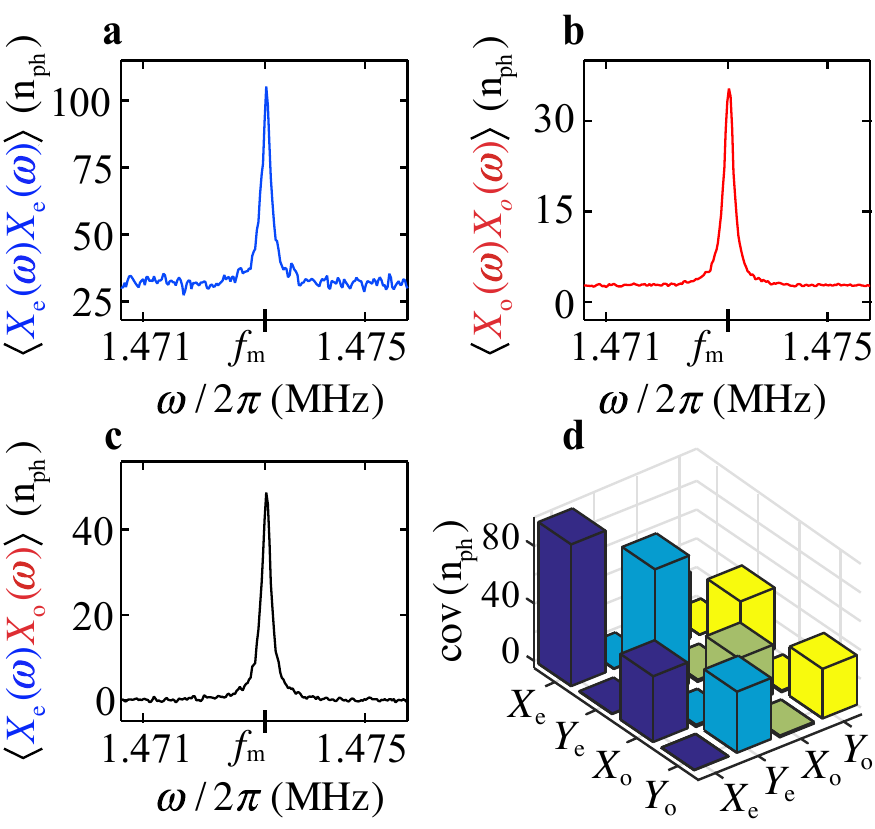}}
\caption{\textbf{Electro-optic correlations. a,} Microwave real spectral density, $\langle X_\mathrm{e}(\omega)X_\mathrm{e}(\omega) \rangle$, in units of photons referred to converter output, showing thermal-mechanical noise peak and background noise from microwave measurement chain.
\textbf{b,} The optical real spectral density, $\langle X_\mathrm{o}(\omega)X_\mathrm{o}(\omega) \rangle$, exhibits similar features.
\textbf{c,} Real cross-spectral density, $\langle X_\mathrm{e}(\omega)X_\mathrm{o}(\omega) \rangle$, illustrating that thermal noise exhibits perfect classical correlations between outputs and background noise is uncorrelated.
\textbf{d,} Covariance matrix obtained from averaging the spectral and cross-spectral densities around $\omega_\mathrm{m}$ with a $2 \pi \times 50~\mathrm{Hz}$ bandwidth, indicating similar behavior for the imaginary microwave and optical spectral densities and the imaginary cross-spectral density; no correlations exist between real and imaginary parts.}
\label{correlations}
\end{minipage}
\end{figure}

The microwave real power spectral density, $\langle X_\mathrm{e}(\omega) X_\mathrm{e}(\omega) \rangle$, exhibits a peak of width $\Gamma_\mathrm{T}/2\pi=200~\mathrm{Hz}$ around $\omega/2\pi=f_\mathrm{m}$ (Fig.~\ref{correlations}a).
The peak height relative to background, $\langle (X^{\mathrm{(th)}}_\mathrm{e})^2\rangle = 69.2~\mathrm{photons}$, is obtained from a Lorentzian fit to the data and attributed to thermally driven mechanical motion at a bath temperature $T = 87 \pm 4 ~\mathrm{mK}$ (see supplement), which exceeds the cryostat base temperature of $35~\mathrm{mK}$.
The elevated membrane temperature is consistent with independently measured optical heating (see supplement).
The background noise level, 31.8 photons, corresponds to $n_\mathrm{e}=29.6~\mathrm{photons}$ from vacuum noise and the added noise of the microwave measurement chain, with the remaining 2.2 photons due primarily to parameter noise in the LC circuit (independently calibrated, see supplement).

The optical real spectral density, $\langle X_\mathrm{o}(\omega) X_\mathrm{o}(\omega) \rangle$, shows a similar peak (Fig.~\ref{correlations}b).
The peak height above noise, $\langle (X^{\mathrm{(th)}}_\mathrm{o})^2\rangle = 33.1~\mathrm{photons}$, provides a second measure of the bath temperature, $T = 80 \pm 4 ~\mathrm{mK}$, consistent with Fig.~\ref{correlations}a.
The background noise, $n_\mathrm{o}=2.7$ photons, corresponds to vacuum noise plus the effect of loss in the optical measurement chain.
Note that $n_\mathrm{o} < n_\mathrm{e}$, indicating that the optical measurement apparatus is closer to ideal.

The real cross-spectral density, $\langle X_\mathrm{e}(\omega) X_\mathrm{o}(\omega)\rangle$, has a $47.7$ photon peak at $\omega/2\pi=f_\mathrm{m}$, indicating that the thermal fluctuations are common to both outputs (Fig.~\ref{correlations}c).
The difference between the observed correlations and their maximum classical value, $\langle X_\mathrm{e}X_\mathrm{o}\rangle - \sqrt{\langle (X^{\mathrm{(th)}}_\mathrm{e})^2\rangle}\sqrt{\langle (X^{\mathrm{(th)}}_\mathrm{o})^2\rangle} = -0.2\pm 0.3~\mathrm{photons}$, is consistent with zero, indicating that thermal noise is perfectly correlated between the two outputs, as expected from optomechanical theory (see supplement).
Away from mechanical resonance the cross-correlation vanishes, as expected for uncorrelated noise from the independent measurement chains. Similar behavior is observed for $\langle Y_\mathrm{e}(\omega) Y_\mathrm{o}(\omega)\rangle$ (Fig.~\ref{correlations}d).

\begin{figure*}
\scalebox{1}{\includegraphics{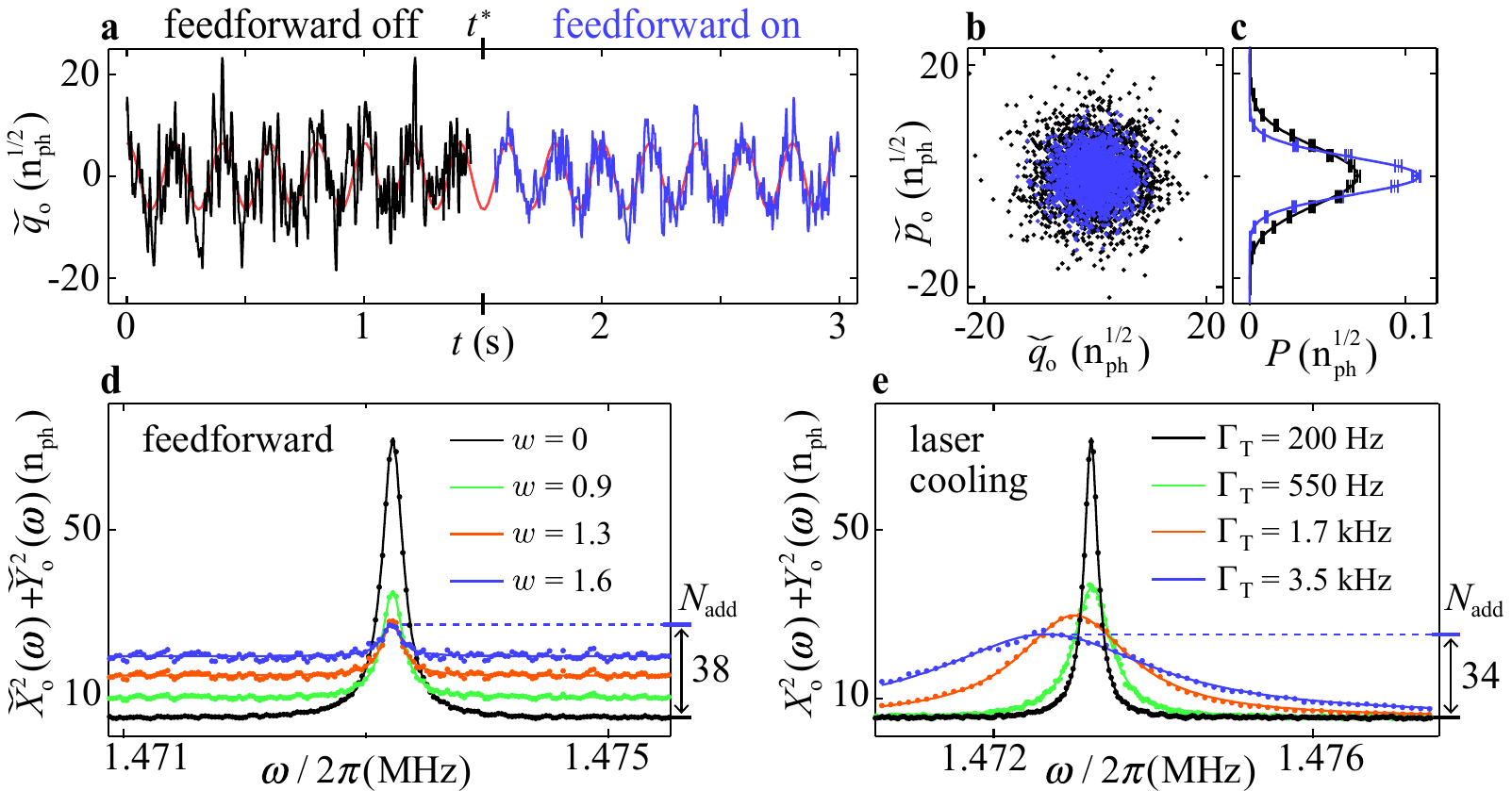}}
\caption{\textbf{Feedforward operation of a microwave-mechanical-optical converter. a,} Fed-forward optical quadrature $\check{q}_\mathrm{o}$ versus time with feedforward off ($t < t^*$) and feedforward on ($t > t^*$). A weak signal incident on the microwave port is recovered at the optical port.
\textbf{b,} Repeated measurements of both optical quadratures $(\check{q}_\mathrm{o},\check{p}_\mathrm{o})$ with and without feedforward. Signal tone turned off.
\textbf{c,} $P(\check{p}_\mathrm{o})$, the inferred probability density of $\check{p}_\mathrm{o}$, with  and without feedforward, with Gaussian fits. The variance decreases by $59\%$ with feedforward on.
Error bars obtained from Poisson counting statistics.
\textbf{d,} Total noise power spectra, $\check{X}_\mathrm{o}^2(\omega) + \check{Y}_\mathrm{o}^2(\omega)$, for different feedforward weights, $w$, with $\Gamma_\mathrm{T}=2 \pi\times200~\mathrm{Hz}$.
Best added noise in photons referred to converter input, $N_\mathrm{add}$, indicated on right axis.
\textbf{e,} Total noise power spectra, $X_\mathrm{o}^2(\omega) + Y_\mathrm{o}^2(\omega)$, for different laser cooling rates, $\Gamma_\mathrm{T}$, while maintaining $\Gamma_\mathrm{e} \approx \Gamma_\mathrm{o}$.
Best added noise in photons referred to converter input, $N_\mathrm{add}$, indicated on right axis.
}
\label{feedforward}
\end{figure*}

Harnessing the observed correlations, we use classical feedforward to recover a weak upconverted signal.
With the weak signal incident on the microwave port and detuned from the pump by $\delta/2\pi = f_\mathrm{m}+5~\mathrm{Hz}$, microwave reflection and converter transmission to the optical port are simultaneously measured.
The microwave reflection is fed forward to remove noise from the upconverted optical signal.
In a quadrature picture (see supplement) where demodulated microwave and optical fields are described in the time domain by $( q(t),p(t) )$ position-momentum pairs, the fed-forward optical quadrature $\check{q}_\mathrm{o}$ is given by
\begin{equation} \label{eq:ffquad}
\check{q}_\mathrm{o} = q_\mathrm{o} - w \sqrt{ \frac{ n_\mathrm{o} }{ n_\mathrm{e} } } q_\mathrm{e},
\end{equation}
where $q_\mathrm{o}$ is the measured optical quadrature, $q_\mathrm{e}$ is the microwave quadrature, and $w$ is the feedforward weight; a similar definition is used for $\check{p}$.

Initially, with feedforward turned off ($w = 0$ for $t < t^*$), the upconverted signal, a $2\pi \times 5~\mathrm{Hz}$ quadrature oscillation, is difficult to resolve (Fig.~\ref{feedforward}a). 
After feedforward is turned on ($w = 1.6$ for $t > t^*$), the weak signal becomes clearly visible.
To quantify the improvement, the optical quadratures $(\check{q}_\mathrm{o}, \check{p}_\mathrm{o})$ are repeatedly measured without a signal tone, as shown in Fig.~\ref{feedforward}b.
The quadratures are Gaussian distributed (Fig.~\ref{feedforward}c), and feedforward reduces each quadrature variance by $59\%$.

Feedforward performance is limited mainly by the addition of uncorrelated noise from imperfect measurement chains.
This limitation can be understood by examining the power spectral density of the feedforward signal, $\langle\check{X}^2_\mathrm{o}\rangle+\langle\check{Y}^2_\mathrm{o}\rangle$, where
\begin{eqnarray}
\langle \check{X}^2_\mathrm{o} \rangle &=& \langle X_\mathrm{o}^2 \rangle + w^2 \frac{n_\mathrm{o}}{n_\mathrm{e}} \langle X_\mathrm{e}^2 \rangle - 2 w \sqrt{\frac{n_\mathrm{o}}{n_\mathrm{e}}} \langle X_\mathrm{e} X_\mathrm{o} \rangle, \\
\langle \check{Y}^2_\mathrm{o} \rangle  &=& \langle Y_\mathrm{o}^2 \rangle + w^2 \frac{n_\mathrm{o}}{n_\mathrm{e}} \langle Y_\mathrm{e}^2 \rangle - 2 w \sqrt{\frac{n_\mathrm{o}}{n_\mathrm{e}}} \langle Y_\mathrm{e} Y_\mathrm{o} \rangle,
\end{eqnarray}
in terms of the measured real and imaginary spectral densities. 
As shown in Fig.~\ref{feedforward}d, as the feedforward weight $w$ is increased, the peak noise power near $f_\mathrm{m}$ decreases due to the presence of correlations near mechanical resonance.
However, Fig.~\ref{feedforward}d also illustrates that the noise power off resonance increases with increasing $w$. At $w=1$, the background noise receives equal contributions from the optical measurement noise and fed-forward noise from the microwave measurement chain; for $w>1$ the noise introduced by feedforward will dominate off-resonance while the thermal noise around $f_m$ continues to decrease. 

For reporting noise performance of the converter, the relevant metric is noise referred to converter input, which can be calculated from the output noise on resonance by dividing by the apparent converter efficiency, $\mathcal{A}\cdot\eta$.
One can then consider an added feedforward noise at converter input (right side of Fig.~\ref{feedforward}d), reflecting how much the observed noise exceeds the amount due to the imperfect measurement chains and vacuum noise.
Feedforward with $w=1.6$ effectively adds $N_\mathrm{add}=38~\mathrm{photons}$ of noise to the converter input, the majority of which is fed-forward microwave measurement noise.

It is interesting to make a comparison between feedforward operation and laser cooling (electro/optomechanical damping) of the mechanical oscillator.
As shown in Fig.~\ref{feedforward}e, increasing the total damping, $\Gamma_\mathrm{T}$, while maintaining matching, $\Gamma_\mathrm{e} \approx \Gamma_\mathrm{o}$, improves noise performance while increasing bandwidth.
Damping rates are limited by laser-induced heating of the superconductor and LC parameter noise (see supplement).
Laser cooling achieved a best value of $N_\mathrm{add} = 34~\mathrm{photons}$ of input-referred added noise  (right side of Fig.~\ref{feedforward}e), comparable to the best feedforward performance.
Unlike feedforward operation, laser cooling does not affect the background noise.

Although feedforward and laser cooling achieve comparable noise performance in the current setup, their limitations are quite different.
In the presence of technical noise that limits damping rates, the performance of laser cooling is set by $n_\mathrm{th,m} \gamma_\mathrm{m}$, where $n_\mathrm{th,m}$ is the thermal phonon occupancy of the membrane.
Indeed, we have studied laser cooling in a different converter with 10 times lower mechanical dissipation but a lower conversion efficiency of $12\%$, and achieved an $N_\mathrm{add} = 13~\mathrm{photons}$ of added noise. With feedforward, on the other hand, one can always completely eliminate the effect of thermal noise at the expense of feeding forward measurement noise.
Provided $\Gamma_\mathrm{e}, \Gamma_\mathrm{o} \gg \gamma_\mathrm{m}$, feedforward performance is determined solely by the measurement apparatus, rather than coupling to the thermal bath. 
It should be emphasized that for any low-frequency mechanical mode this is a much less stringent condition on damping rates than high cooperativity ($\Gamma_\mathrm{e}, \Gamma_\mathrm{o} \gg n_\mathrm{th,m}\gamma_\mathrm{m}$).
For example, in a 4~K experiment with a similar measurement setup, one would expect feedforward performance similar to that observed here, but laser cooling to be orders of magnitude less effective.

Looking ahead to quantum feedforward operation, a central challenge is improving microwave and optical measurement performance.  In particular the microwave measurement noise could be improved by using a quantum-limited microwave amplifier~\cite{teufel_nanomechanical_2009}.
Meanwhile, more theoretical work is needed to thoroughly explore the electro-optomechanical correlations identified here, and to study optimal qubit encodings for microwave-optical conversion with feedforward.

\subsection*{Methods}
\subsubsection*{Device parameters}
We realize an electro-optic converter by simultaneously coupling microwave and optical resonators to a single mode of a mechanical oscillator. The mechanical oscillator is a thin,  suspended dielectric membrane,  with resonant frequency $\omega_\mathrm{m} / 2 \pi = f_\mathrm{m} = 1.473~\mathrm{MHz}$ for the mode of interest. A portion of the membrane is metallized and arranged as a mechanically compliant capacitor in a superconducting LC circuit, with resonant frequency $\omega_\mathrm{e}/2\pi = 6.16~\mathrm{GHz}$ and linewidth $\kappa_\mathrm{e} = 2 \pi \times 2.5~\mathrm{MHz}$.
Another portion of the membrane is situated in the mode of a Fabry-Perot optical cavity with resonant frequency $\omega_\mathrm{o}/2\pi = 281.8~\mathrm{THz}$ and linewidth $\kappa_\mathrm{o} = 2 \pi \times 2.1~\mathrm{MHz}$.
Vibrational motion of the membrane modulates $\omega_\mathrm{e}$ by $G_\mathrm{e} \approx 2\pi\times8~\text{MHz/nm}$ and $\omega_\mathrm{o}$ by $G_\mathrm{o} \approx 2\pi\times13~\text{MHz/nm}$.
The entire assembly is housed in a cryostat with base temperature $T=35~\mathrm{mK}$~\cite{peterson_laser_2016}, where both microwave and optical modes are close to their quantum ground state.

With a strong red-detuned pump beam incident on the optical cavity, the optomechanical interaction couples the mechanical oscillator to propagating optical fields at a rate $\Gamma_\mathrm{o}$ that exceeds the intrinsic mechanical damping rate $\gamma_\mathrm{m} = 2\pi\times11~\mathrm{Hz}$. At pump detuning $\Delta_\mathrm{o} = -\omega_\mathrm{m}$ and in the resolved sideband limit ($4\omega_\mathrm{m}/\kappa_\mathrm{o} \gg 1$), the optomechanical damping rate has the simple form\cite{andrews_bidirectional_2014}
\begin{equation}
\Gamma_\mathrm{o} = \frac{4G_\mathrm{o}^2x_\mathrm{zp}^2\alpha^2_\mathrm{o}}{\kappa_\mathrm{o}},
\end{equation}
where $\alpha_\mathrm{o}$ is the intracavity pump amplitude and $x_\mathrm{zp}$ is the zero-point amplitude of the membrane mode. The electromechancial damping in the presence of a strong red-detuned microwave pump tone has the same form:
\begin{equation}
\Gamma_\mathrm{e} = \frac{4G_\mathrm{e}^2x_\mathrm{zp}^2\alpha^2_\mathrm{e}}{\kappa_\mathrm{e}}.
\end{equation}
In practice, our converter is moderately sideband-resolved ($4\omega_\mathrm{m}/\kappa_\mathrm{o} \approx 4\omega_\mathrm{m}/\kappa_\mathrm{e} \approx 2.5$), with pump detunings $\Delta_\mathrm{e} / 2 \pi =-1.47~\mathrm{MHz}$ and $\Delta_\mathrm{o}/2 \pi = -1.11~\mathrm{MHz}$. These details change the precise form of the expressions for $\Gamma_\mathrm{o}$ and $\Gamma_\mathrm{e}$ but not the essential physics.

We set the optical pump detuning $\Delta_\mathrm{o}$ to maximize the optomechanical damping rate per incident photon. In the resolved sideband limit, this maximum damping occurs at $\Delta_\mathrm{o} = -\omega_\mathrm{m}$, but this is not generally the case with imperfect sideband resolution.

\subsubsection*{Fabrication Details}
The mechanical oscillator is a $100~\mathrm{nm}$ thick, $500~\mathrm{\mu m}$ wide silicon nitride membrane suspended from a silicon chip. A $25~\mathrm{nm}$ thick niobium film which serves as one capacitor pad is fabricated on one quadrant of the membrane before it is released. The membrane chip is flipped over and affixed to a second silicon chip, on which a microfabricated niobium circuit comprising an inductor and a second capacitor pad was previously patterned using standard lithographic techniques. The full fabrication process has been described elsewhere \cite{andrews_2015}.
The flip-chip assembly is constructed with a West Bond manual die bonder, and the chips are affixed using Stycast 2850.
In the fully assembled flip-chip device, the two niobium pads form a parallel-plate capacitor with a plate spacing of $300~\mathrm{nm}$, and the  $6.16~\mathrm{GHz}$ resonant frequency of the resulting LC circuit is modulated by vibrations of mode of interest.
Propagating microwaves are wirelessly coupled to the LC circuit through a re-entrant microwave cavity, which also holds mirrors that form a Fabry-Perot optical cavity with a $281.8~\mathrm{THz}$ resonant frequency \cite{menke_reconfigurable_2017}.
The optical cavity comprises two mirrors, with 29 ppm and 98 ppm power transmission respectively, separated by $2.6~\mathrm{mm}$.
The chip assembly is placed in the standing wave of the optical cavity with the membrane $750~\mathrm{\mu m}$ from the high-transmission mirror, such that the membrane's vibrations modulate the cavity's resonant frequency \cite{thompson_strong_2008}. The optical mode intersects the membrane in the quadrant opposite the capacitor, and these two regions of the membrane move in phase for the mechanical mode of interest.

\subsubsection*{Data Availability}
The data that support the plots within this paper and other findings of this study are available from the corresponding author upon reasonable request.

\acknowledgements{We thank James Thompson and Murray Holland for fruitful conversations and Katarina Cicak for assistance with device fabrication.  We acknowledge funding from AFOSR MURI grant number FA9550-15-1-0015, NSF under grant number PHYS 1734006, DURIP, and AFOSR PECASE.}

\subsection*{Author Contributions}
A.P.H., P.S.B., and M.D.U.~conducted the experiment and analyzed data. A.P.H, P.S.B., M.D.U., R.W.P., and N.S.K.~designed and constructed the measurement network. M.D.U. and R.W.P.~designed and constructed the optical cavity. P.S.B.~designed and fabricated the flip-chip device. A.P.H.~and G.S.~developed feedforward theory. A.P.H., P.S.B., M.D.U., B.M.B., G.S., K.W.L., and C.A.R.~wrote the manuscript. C.A.R.~and K.W.L.~supervised the work. All authors commented on the results and manuscript.

\subsection*{Competing Financial Interests statement}
The authors declare no competing financial interests. 

\bibliographystyle{naturemag_thesiscase} 

\onecolumngrid


\newpage


\makeatletter 
\renewcommand{\thefigure}{S\@arabic\c@figure}
\renewcommand{\thetable}{S\arabic{table}}
\renewcommand{\theequation}{S\arabic{equation}}
\makeatother
\setcounter{figure}{0}    

\setcounter{table}{0}    

\setcounter{equation}{0}

\section*{Supplementary Information}

%

\section{System parameter summary}

System parameters are summarized in Table~\ref{tab:params} for the configuration with $\eta_\mathrm{M}=0.43$.
These parameters apply to all data in the main text, except Fig.~2c, where $\eta_\mathrm{M}$ is varied.

\begin{table}
\caption{\label{tab:params}Converter parameters for $\eta_\mathrm{M}=0.43$ configuration. }
\begin{ruledtabular}
\begin{tabular}{c c c}
$\kappa_{\text{e}}/2\pi$ 		& 2.5						& MHz\\
$\kappa_{\text{ex,e}}/2\pi$ 		& 2.3						& MHz\\
$\kappa_{\text{int,e}}/2\pi$ 		& 0.2						& MHz\\
$\Delta_{\text{e}}/2\pi$			& -1.47				& MHz\\
$G_{\text{e}}x_\mathrm{zp}/2\pi$ 	& 3.8	& Hz\\
$\kappa_{\text{o}}/2\pi$ 			& 2.1						& MHz\\
$\kappa_{\text{ex,o}}/2\pi$ 		& 1.1						& MHz\\
$\epsilon$        & 0.87                      & \\
$\kappa_{\text{B,o}}/2\pi + \kappa_{\text{int,o}}/2\pi$ 		& 1.0						& MHz\\
$\Delta_{\text{o}}/2\pi$ 			& -1.11						& MHz\\
$G_{\text{o}}x_\mathrm{zp}/2\pi$ 	& 6.6			& Hz\\
$\omega_{\text{m}}/2\pi = f_\mathrm{m}$ 		& 1.4732					& MHz\\
$\gamma_{\text{m}}/2\pi$ 		& 11	& Hz
\end{tabular}
\end{ruledtabular}
\end{table}

\section{Measurement network diagram and calibrations}
\begin{figure}
\begin{minipage}{\linewidth}
\scalebox{.63}{\includegraphics{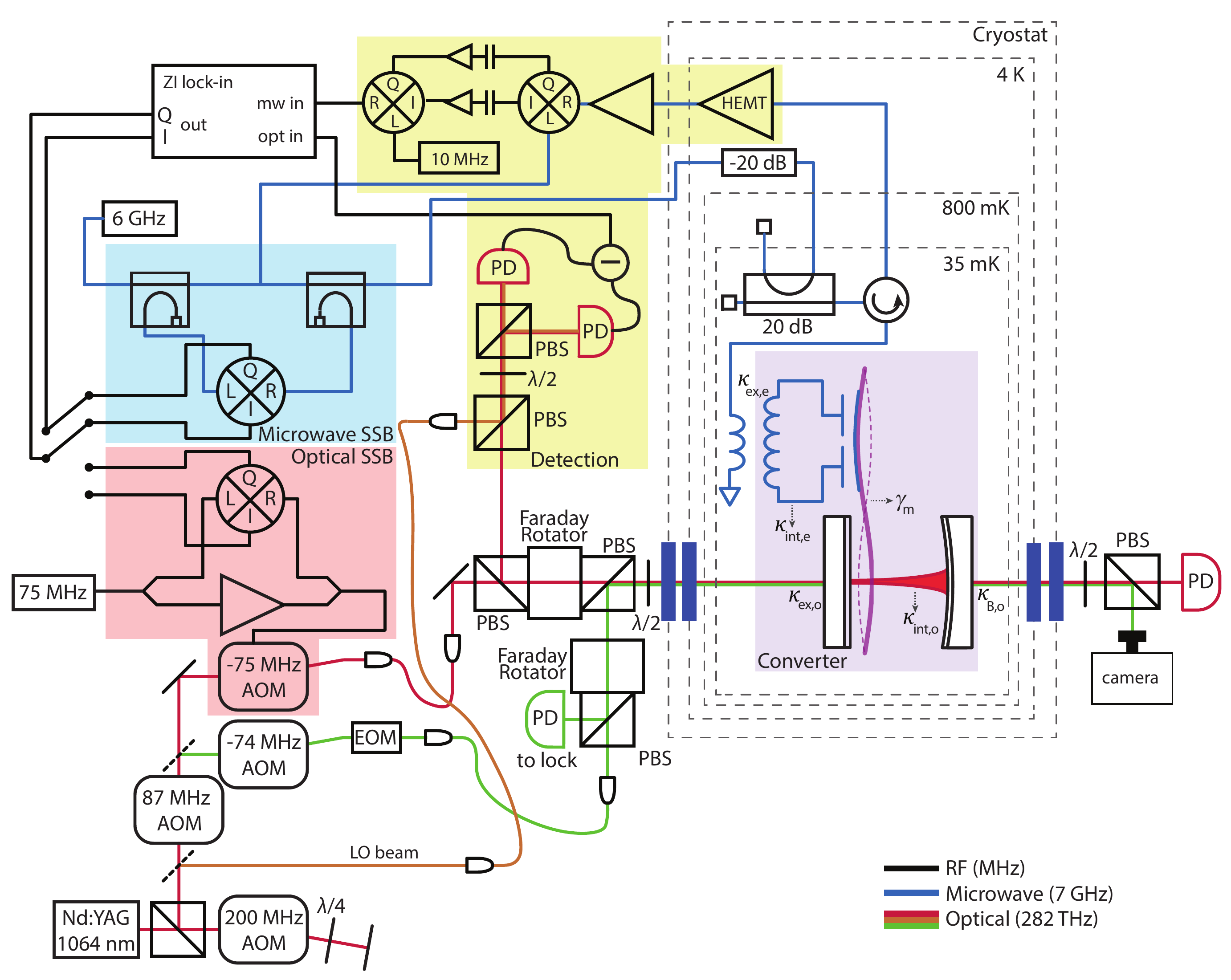}}
\caption{Measurement Network. 
A Zurich Instruments Lock-in Amplifier (ZI) is used as a spectrum and network analyzer. Outputs are driven $90^\circ$ out of phase to produce a single sideband probe tone. 
In the microwave single sideband generator (blue box), a mixer is used to produce a single upper sideband directly on the microwave pump. 
The optical single sideband modulation (red box) is first mixed with an intermediate $75~\mathrm{MHz}$ tone to drive an AOM. 
A lock beam (green) is used to reference and lock the optical cavity. 
The converter (purple box) is thermalized to the base plate of an optical access dilution refrigerator (grey dashed box), and microwave and optical tones are routed in with appropriate filtering.  
A directional coupler is used at base to provide filtering and an effective cold load. 
Transmission through the optical cavity can be used for characterization. 
Reflected and converted signals are routed by circulators to detection (yellow). 
A cryogenic HEMT amplifier is used for microwave amplification. 
The microwave output signal is demodulated with an I/Q mixer, remodulated at a frequency of $10~\mathrm{MHz}$, and then routed to the ZI. 
The optical beam is mixed with a local oscillator beam (orange) and measured in heterodyne. 
Spectrum analyzer measurements can be made by turning the ZI outputs off and fixing the demodulation frequency.}
\label{diagram}
\end{minipage}
\end{figure}

Fig. \ref{diagram} shows a detailed schematic of the measurement network. 


\subsection{Data acquistion}
Data acquisition is performed with a Zurich Instruments HF2LI lock-in amplifier.
Scattering parameters are measured by sweeping an output tone and demodulator frequency, and spectra can be measured by taking a time trace and computing the Fourier transform. 
The converter is operated at the base temperature of a dilution refrigerator with optical access, and a cryogenic HEMT amplifier is used for microwave measurements. 
Microwave signals are measured by demodulating with an I/Q mixer, remodulating at a frequency of 10 MHz, then feeding into the HF2LI.
The demodulate-remodulate procedure allows both microwave quadratures to be measured by a single physical channel, and makes the signal processing functions performed by the HF2LI symmetric between the microwave and optical domains.
Optical signals are measured by combining the beam exiting the cavity with an LO on a beam splitter and reading out the difference port of a balanced heterodyne detector.


When computing spectra, an incident signal $s(t)$ is demodulated at a frequency $f_\mathrm{d} = f_\mathrm{c} + f_\mathrm{m}$, where $f_\mathrm{c}$ is the carrier frequency: $f_\mathrm{c}=10~\mathrm{MHz}$ for microwave measurements and $f_\mathrm{c}=12~\mathrm{MHz}$ for optical measurements.
The resulting quadratures, $q(t)$ and $p(t)$, are used to compute, 
\begin{equation}
a(t) = ( q(t) + i p(t) ) e^{i 2 \pi f_\mathrm{d} t},
\end{equation}
which is the complex-equivalent signal of $s(t)$, 
\begin{equation}
a(t) = s(t) + i H[ s( t ) ],
\end{equation}
where $H$ denotes the Hilbert transform.
The Fourier transform of $a(t)$, $a( \omega ) = \int_{-\infty}^{+\infty} a(t) e^{-i \omega t}\, \mathrm{d}t$ is then computed approximately using  DFT.
$a( \omega )$ is a complex-valued function, 
\begin{equation}
a(\omega) = X( \omega ) + i Y( \omega ),
\end{equation}
with real component $X(\omega)$ and imaginary component $Y(\omega)$, and is zero when $\omega < 0$.
In the limit of narrowband demodulation the quadrature variances satisfy e.g., $\mathrm{Var}(q_\mathrm{o}) =  \langle X_\mathrm{o}( \omega_\mathrm{m} ) X_\mathrm{o}( \omega_\mathrm{m} ) \rangle$.
A demodulation bandwidth of $2 \pi \times 50~\mathrm{Hz}$ is used for all quadrature data.
Note that away from mechanical resonance, $X(\omega)$ is not just the Fourier transform of $q(t)$, but rather contains contributions from both $q(t)$ and $p(t)$.
In the main text, spectra are displayed shifted so that $f_\mathrm{c}$ appears at zero frequency.


\subsection{Microwave and optical calibrations} \label{sec:omcal}
\begin{figure}
\begin{minipage}{\linewidth}
\scalebox{.9}{\includegraphics{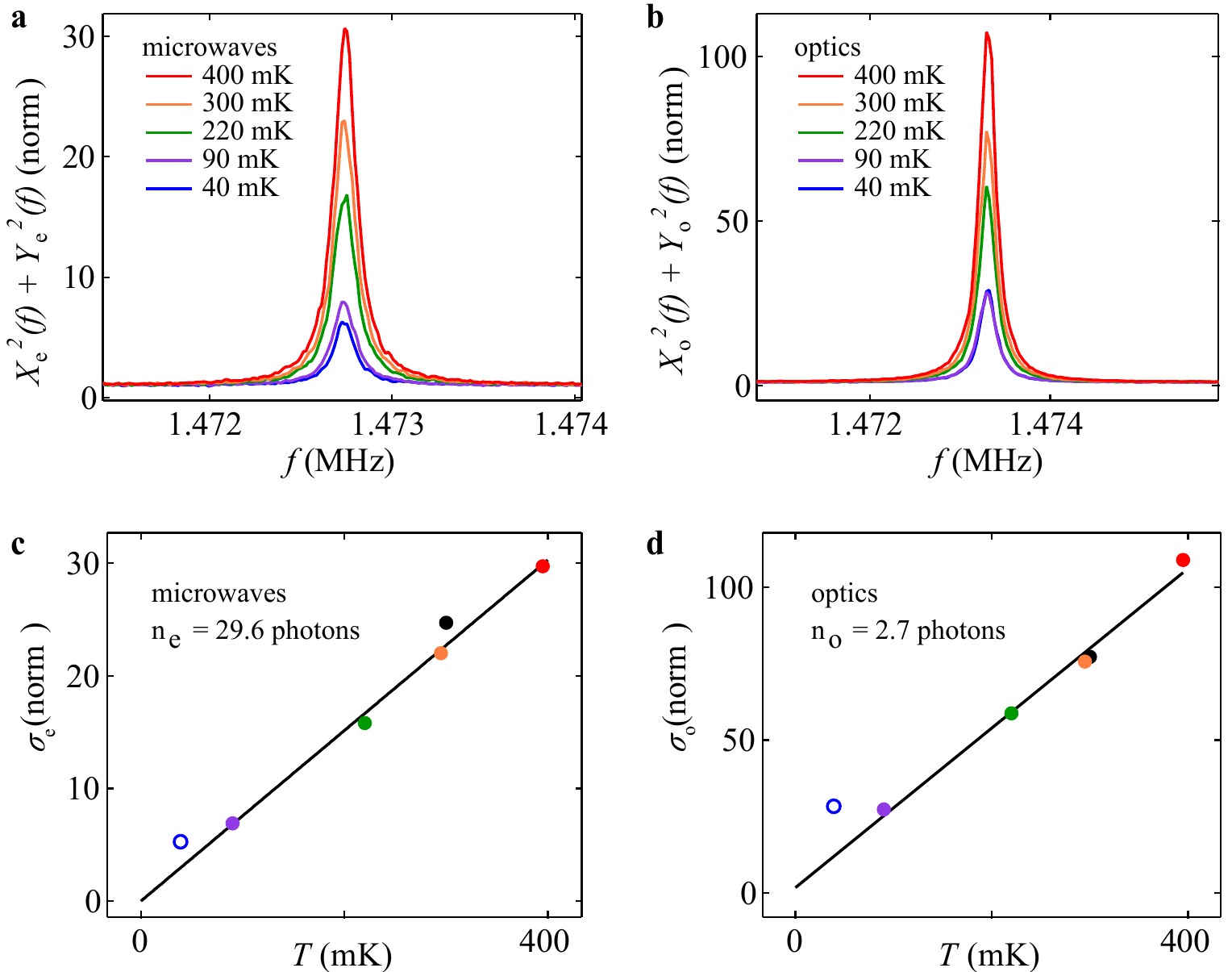}}
\caption{(a) Total microwave noise power spectra, $X_\mathrm{e}^2(f) + Y_\mathrm{e}^2(f)$, shown for a range of base plate temperatures. 
Optical pump off. 
Noise is normalized to unity with pumps off. 
(b) Total optical noise power spectra, $X_\mathrm{o}^2(f) + Y_\mathrm{o}^2(f)$, shown for a range of base plate temperatures, with the same normalization procedure.
Microwave pump off.
(c) Microwave peak height $\sigma_e$ vs.~temperature $T$, with a single-parameter linear fit to extract total noise (see text). 
Open data point, indicating imperfect thermalization to the cryostat at low temperature, is excluded from fit.
(d) Optical peak height $\sigma_o$ vs.~temperature $T$, with a similar fit. 
Open data point, indicating imperfect thermalization to the cryostat at low temperature, is excluded from fit.}
\label{fig:noise}
\end{minipage}
\end{figure}

A temperature sweep of the base plate of the dilution refrigerator is used to calibrate output-referred noise spectra like those shown in Fig.~3 of the main text. Raw spectra are normalized to the noise level with the pump off, to which the only contributions are vacuum noise and measurement chain added noise; the peak height is $\sigma$ in these units.
The peak exhibits a linear temperature dependence due to the thermally driven motion of the mechanical oscillator (Fig.~\ref{fig:noise}), except at the lowest temperatures where the membrane evidently falls out of equilibrium with the dilution refrigerator (open circles, Fig.~\ref{fig:noise}c and d).
The temperature below which the membrane falls out of equilibrium is higher in the optical measurement, indicating that optical heating is a significant factor below $T=100~\mathrm{mK}$.
We find that this membrane heating occurs whenever the locking beam is introduced, but is independent of damping power, consistent with the observations in Ref.~\cite{peterson_laser_2016}.
The spectra are calibrated by fitting the observed peak heights $\sigma(T)$ to an expected linear dependence obtained from combining the full optomechanical equations of motion~\cite{andrews_bidirectional_2014} with independently measured parameters in Table~\ref{tab:params} \cite{purdy_cavity_2012}.
The only fit parameter is a proportionality factor that converts uncalibrated thermal peaks to units of photons at the converter output; the normalization of the uncalibrated spectra implies that the single-quadrature background noise (vacuum noise plus measurement chain added noise) is simply given by the calibration factor for each measurement. This procedure yields a single-quadrature background noise $n_\mathrm{e} = 29.6~\mathrm{photons}$ at the microwave output port and $n_\mathrm{o} = 2.7~\mathrm{photons}$ at the optical output port. 
Inverting this calibration procedure, we can associate the number of output-referred photons on resonance in spectra like those in Fig.~3 of the main text with an effective mechanical bath temperature.

The microwave cavity coupling ratio $\kappa_\mathrm{ex,e}/\kappa_\mathrm{e}$ is easily measured using the HF2LI as a network analyzer (see Sec.~\ref{microwaves}). The optical cavity parameters are needed to predict the optical field output noise using optomechanical theory~\cite{aspelmeyer_cavity_2014}.
The mode matching between the cavity and the pump beam is obtained by directing a transmitted beam and the reflected pump onto the heterodyne detector, and measuring the visibility of the resulting interference pattern (see Sec.~\ref{modematch}).
The optical cavity linewidth is measured by sweeping a probe with two sidebands at a known frequency through resonance.
The optical cavity coupling ratio $\kappa_\mathrm{ex,o}/\kappa_\mathrm{o}$, which is used to determine $\kappa_\mathrm{ex,o}$ in Table~\ref{tab:params}, is then fixed based on the measured converter efficiency.
As a check, $\kappa_\mathrm{ex,o}/\kappa_\mathrm{o}$ is independently extracted from reflection and transmission measurements of both ports of the optical cavity \cite{hood_characterization_2001}.
This procedure yields $\kappa_\mathrm{ex,o}/\kappa_\mathrm{o}=0.53$, which can also be used to determine the expected microwave-optical conversion efficiency, $\epsilon (\kappa_\mathrm{ex,o}/\kappa_\mathrm{o})( \kappa_\mathrm{ex,e}/\kappa_\mathrm{e}) = 43 \pm 4\%$, as reported in the main text.

\subsection{Calibration checks}

As a check of the optical calibration in Sec.~\ref{sec:omcal}, we independently measure the optical path loss and heterodyne detector dark noise.
This measurement gives an effective transmission of $16\%$, or equivalently an expected single-quadrature background noise of $(1/4\times 2) \times 1 / 0.16 = 3.1~\mathrm{photons}$, where the three factors correspond to the single-quadrature vacuum variance, inefficiency of an ideal heterodyne detector, and effective loss, respectively. 
This is in reasonable agreement with the thermal calibration.

\begin{figure}
\begin{minipage}{\linewidth}
\scalebox{.9}{\includegraphics{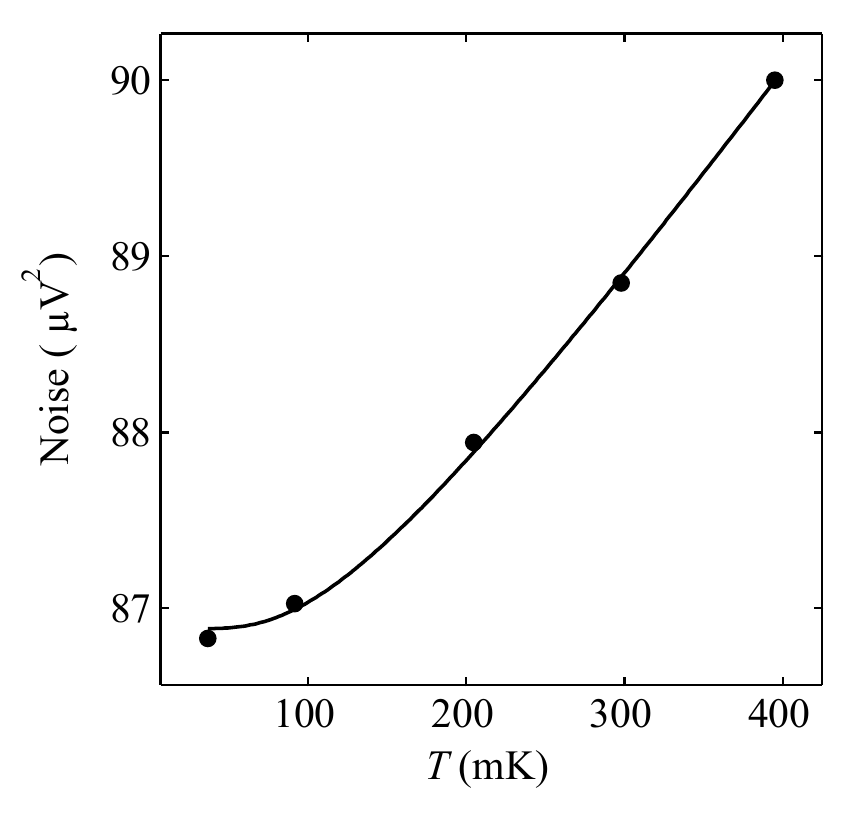}}
\caption{Noise measured in the microwave domain off LC resonance as temperature is varied, fit to Eq.~\ref{eq:hemtnoise}.
}
\label{fig:hemt}
\end{minipage}
\end{figure}

As a check of the microwave calibration in Sec.~\ref{sec:omcal}, the amplifier chain can be characterized by measuring off-resonance noise, as shown in Fig.~\ref{fig:hemt}.
Data are fit to 
\begin{equation} \label{eq:hemtnoise}
S_\mathrm{out} = G\left(\frac{1}{2}\coth\left(\frac{hf}{2k_\mathrm{B}T}\right)+N_\mathrm{HEMT}\right),
\end{equation}
where the two fit parameters are the gain $G$ and two-quadrature receiver added noise $N_\mathrm{HEMT}$, which we expect is dominated by the added noise of the HEMT preamplifier.
We obtain $N_\mathrm{HEMT}=20~\mathrm{photons}$, which implies 10 photons of added noise in a single-quadrature measurement. Comparing this result to the single-quadrature noise in Sec.~\ref{sec:omcal}, $n_\mathrm{e}=29.6~\mathrm{photons}$, naively implies $4.6~\mathrm{dB}$ of loss from the LC circuit output to the effective cold load, a somewhat high value.
This loss inference should be treated with caution, as the temperature sweep in Fig.~\ref{fig:hemt} was performed with some stages of the dilution refrigerator slightly colder, which lowers total chain noise compared to Fig.~\ref{fig:noise}.

With the strong pump powers needed for our experiment, two distinct effects lead to excess noise within the LC circuit bandwidth. First, we observe excess phase noise on resonance which scales with the microwave pump power and overwhelms the intrinsic phase noise of the pump generator. Similar ``parameter noise'' has been observed in other superconducting resonator geometries~\cite{gao_noise_2007}, and may arise from fluctuating two-level systems at the substrate-metal interface or in a surface oxide layer~\cite{gao_experimental_2008}. From independent measurements, we expect parameter noise to contribute 1 photon to a single-quadrature microwave measurement with $\Gamma_\mathrm{e}=2\pi \times 95~\mathrm{Hz}$, as in Fig.~3 of the main text. 
We also observe excess noise in the LC circuit that scales with the total laser power incident on the membrane, to which both the lock beam and the optical pump contribute. From independent measurements, we expect laser heating to contribute an additional 0.5 photons of noise in a single-quadrature microwave measurement with $\Gamma_\mathrm{o}=2\pi \times 95~\mathrm{Hz}$, as in the main text.
The sum of the parameter noise and laser heating contributions is in reasonable agreement with the 2.2 excess photons measured at the microwave output port in Fig.~3a of the main text. 
The effects of LC parameter noise in particular are significantly exacerbated by the much higher pump powers required for laser cooling. 
For example, in the configuration with $\Gamma_\mathrm{T}=2\pi\times3.5~\mathrm{kHz}$ for which we obtained the best laser cooling performance, we attribute 25 photons of added noise referred to converter input to parameter noise in the microwave resonator.

Laser cooling performance is usually quantified via the minimum phonon occupancy $n_\mathrm{f,m}$ of the mechanical resonator, a quantity closely related to the input-referred added noise cited in the main text.  
Laser cooling with $\Gamma_\mathrm{T}/2\pi=3.5~\mathrm{kHz}$ reduced the membrane's phonon occupancy from $n_\mathrm{th,m}\approx1200$ to a best value of $n_\mathrm{f,m}=18.6$.
In the device with 10 times lower mechanical dissipation mentioned in the main text, we obtained a final membrane occupancy of $n_\mathrm{f,m}=4.9$ phonons.

\subsection{Mode matching and efficiency calibration}\label{modematch}

Care must be taken to correctly treat the various optical mode matching factors that impact the probe-tone scattering measurements. We parameterize the mode-matching between two optical fields of the same polarization $E_1$, $E_2$ with the factor 
\begin{equation}
    \epsilon = \frac{\left|{\int E_1^* E_2}\mathrm{d}^2x\right|}{\sqrt{\int\left|E_1\right|^2\mathrm{d}^2x\int\left|E_2\right|^2\mathrm{d}^2x}}.
\end{equation}
Operationally, we direct two beams at different frequencies onto a photodetector; $\epsilon$ is measured as the square root of the fringe visibility.

The converter is calibrated by dividing the transmission in each direction by the off-resonant reflection off each port \cite{andrews_bidirectional_2014}.
In this way the individual gains and losses of the measurement network need not be known, and yet they can still be calibrated out. 
With this calibration procedure, mode matching $\epsilon_\mathrm{d}$ between the incident beam and the cavity mode is automatically included in the downconversion efficiency, and mode matching $\epsilon_\mathrm{u}$ between the outgoing cavity mode and the LO is included in the upconversion efficiency. In practice $\epsilon_\mathrm{u} \approx \epsilon_\mathrm{d}$, and we define the mode-matching factor $\epsilon$ that appears in the expression for the bidirectional conversion efficiency $\eta_M$ in the main text as $\epsilon = \sqrt{\epsilon_\mathrm{u}\epsilon_\mathrm{d}}$.

With a non-ideal optical heterodyne measurement, we must also account for imperfect mode-matching $\epsilon_\mathrm{LO}$ between the local oscillator and the reflection from the optical port, which only affects the off-resonant reflection measurement used to calibrate out path losses.
Since $\epsilon_\mathrm{LO}$ is not an inherent quality of the converter, it should not be included in the measured conversion efficiency, and must be calibrated out separately.
An independent measurement of the interference visibility between the LO and a far-detuned beam reflected off the cavity yields $\epsilon_\mathrm{LO}=0.83$, which is similar to $\epsilon=0.87$ as expected.


\section{Optical cavity linewidth theory}
Both optical cavity mirrors can be translated longitudinally \textit{in situ} via piezoelectric actuators. During operations, one of these actuators is used in a Pound-Drever-Hall cavity locking scheme to keep the optical pump at a fixed detuning from the cavity resonance; the other may be used to change the membrane position within the optical standing wave of the cavity. The membrane and fixed mirror effectively constitute a low-finesse etalon whose transmission depends on their separation. Thus, the optical cavity parameters can be adjusted \textit{in situ} by actuating the fixed mirror piezo. 

\begin{figure}
\begin{minipage}{\linewidth}
\scalebox{.9}{\includegraphics{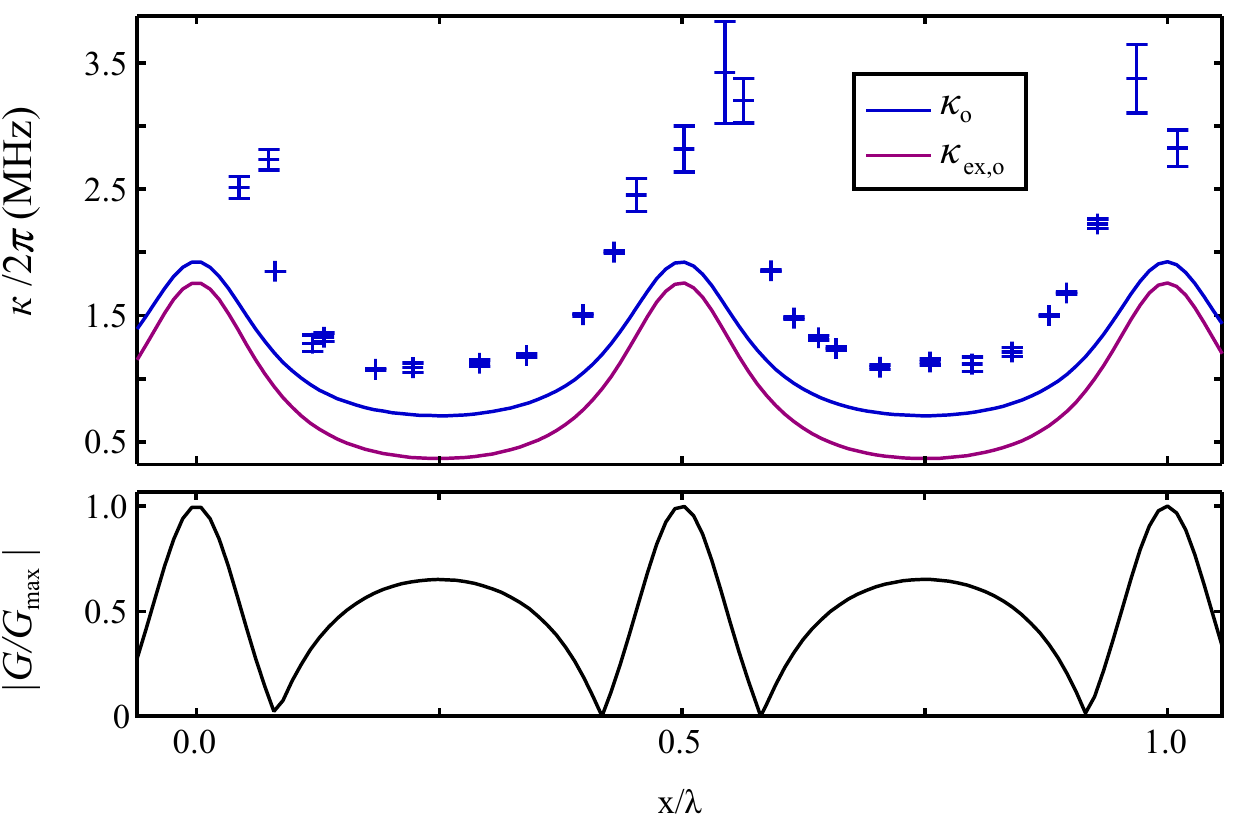}}
\caption{Upper plot: simulated and measured optical cavity linewidth, and simulated external coupling rate as membrane distance from mirror, $750 \rm{~\mu m} + x$, is varied.
Error bars represent the standard deviation of several repeated linewidth measurements. 
Lower plot: absolute value of theoretical optomechanical coupling normalized to maximum coupling.}
\label{opt}
\end{minipage}
\end{figure}

Figure \ref{opt} shows the theoretical and measured optical linewidth and the theoretical optomechanical coupling rate as a function of membrane-mirror separation. 
As the membrane position is varied, the light energy is stored predominately to one side or the other of the membrane \cite{purdy_cavity_2012}, and the external coupling factors associated with the high-throughput input mirror and the low-throughput back mirror ($\kappa_{\text{ex,o}}$ and $\kappa_{\text{B,o}}$, respectively) vary accordingly. 
In order to further increase $\kappa_{\text{ex,o}}/\kappa_\text{o}$, the optical cavity mirrors have unequal transmission rates. 
This cavity asymmetry results in variation in the total optical cavity linewidth $\kappa_\text{o}$ as the membrane position is varied. 
Furthermore, our membrane is placed closer to the input mirror, resulting in enhanced optomechanical coupling $G_\mathrm{o}$ when $\kappa_{\text{ex,o}}/\kappa_\text{o}$ is maximized.
Ideally, we would operate at a position where $G_\mathrm{o}$ and $\kappa_{\mathrm{ex,o}}$ are simultaneously maximized.
However, due to optical instability, a configuration with $\eta_\text{M}=0.43$ and $\kappa_\mathrm{ex,o}/\kappa_\mathrm{o}=0.54$ was used instead for noise measurements.

\section{Optomechanical response theory}
Reflection measurements of the optical cavity can be understood using the mechanically mediated state transfer defined in the supplement of Ref.~\cite{andrews_bidirectional_2014}.
To our knowledge, electromechanically induced optical absorption has not been previously reported, and is therefore examined in more detail here.
The optical cavity has a Lorentzian response and shows up as dip in reflection, from which we detune a pump tone and study mechanical response.

\begin{figure}
\begin{minipage}{\linewidth}
\scalebox{.9}{\includegraphics{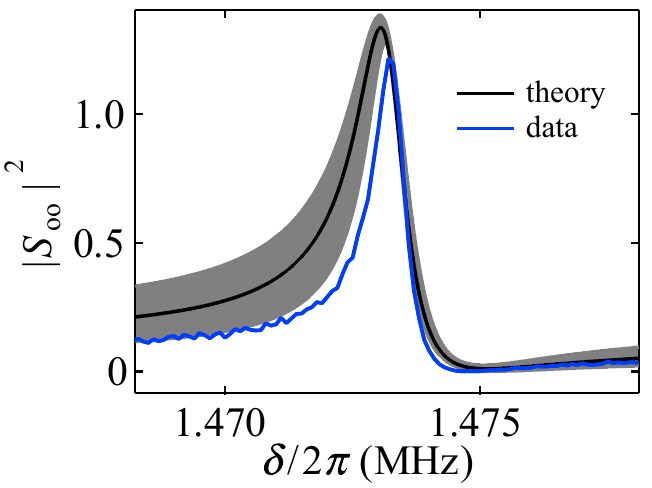}}
\caption{Measured scattering parameter of the optical port vs probe frequency $\delta$ (blue). 
Black line shows expected response from independent parameter measurements. 
Grey band expresses uncertainty of system parameters.}
\label{fig:s22}
\end{minipage}
\end{figure}

Fig.~\ref{fig:s22} shows narrow-band measured and expected scattering parameters of the mechanical response for the optical port, with $\Gamma_\text{e}\ll\Gamma_\text{o}$.
Fixing all the cavity and pump parameters in the equations of motion generates a line shape that reproduces salient features observed in the experiment. The peak value exceeds one due to gain from imperfect sideband resolution, and the Fano-like shape arises from the mechanical response not being on the flat bottom of the dip, but on the slope of the optical cavity response (due to $\Delta_\mathrm{o} \neq -\omega_\mathrm{m}$; see methods).

\section{Microwave cavity reflection}\label{microwaves}
The microwave network is characterized by reflection scattering measurements.
Off resonance of the LC circuit, all the power is reflected, so off-resonance measurements therefore measure the net loss/gain of the microwave chain. 
By normalizing to the off-resonance value we can isolate the response of the LC cavity.
The transfer function of a one-port cavity is 
\begin{equation} \label{s11}
S(\Delta)=-\frac{2i\Delta + \kappa_\mathrm{int}-\kappa_\mathrm{ex}}{2i\Delta + \kappa_\mathrm{int}+\kappa_\mathrm{ex}},
\end{equation}
where $\Delta$ is the detuning between the probe and cavity, $\kappa_\mathrm{int}$ is the internal loss rate, and $\kappa_\mathrm{ex}$ is the external coupling rate.

Using the amplitude and phase of the transfer function, we can extract internal and external coupling rates, $\kappa_\mathrm{ex,e}/2\pi = 2.3~ \mathrm{MHz}$ and $\kappa_\mathrm{int,e}/2\pi = 0.2~ \mathrm{MHz}$.
The background level in the narrowband microwave reflection measurement shown in Fig.~2 of the main text matches the resonant dip value
\begin{equation} 
|S(0)|^2=0.69.
\end{equation}

\begin{figure}
\begin{minipage}{\linewidth}
\scalebox{.9}{\includegraphics{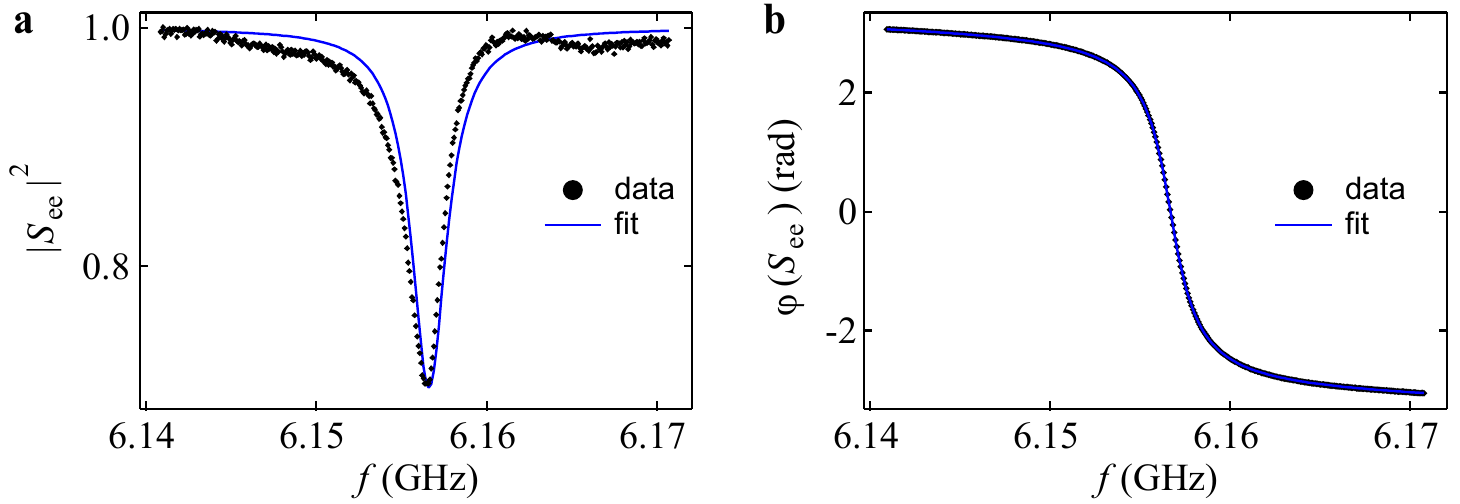}}
\caption{Measured scattering parameter of microwave circuit (black), with fit (blue) to extract cavity coupling rates. 
(a) Power response shows Lorentzian dip with some background ripple characteristic of wide band measurement. 
(b) Phase response.}
\label{mwcoupling}
\end{minipage}
\end{figure}

\section{Electro-optic correlation theory} \label{sec:corrtheory}
Spectra of the output fields are expressed in terms of a matrix $\mathbf{C}( \omega )$, whose elements are given by
\begin{equation}
C_{ij}( \omega ) \delta( \omega - \omega' ) = \frac{1}{2} \langle \{ [ \mathbf{a}_{\mathrm{out},i} (\omega') ]^\dagger, \mathbf{a}_{\mathrm{out},j}( \omega ) \} \rangle,
\end{equation}
where $\mathbf{a}_\mathrm{out} = ( \hat{ a }_\mathrm{out,B}, \hat{ a }_\mathrm{out,F}, \hat{ b }_\mathrm{out}, \hat{ a }^\dagger_\mathrm{out,B}, \hat{ a }^\dagger_\mathrm{out,F}, \hat{ b }^\dagger_\mathrm{out} )^T$ 
is a vector of optical ($\hat{a}_\mathrm{out}$) and microwave ($\hat{b}_\mathrm{out}$) output field operators.\footnote{Optical operators are written for both the front, $\hat{a}_\mathrm{out,F}$, and back, $\hat{a}_\mathrm{out,B}$, optical ports, but in the main text the front port is simply referred to as `external,' as it is the only port where signals are measured.}
Output fields are related to input fields by the matrix $\mathbf{\Xi}( \omega )$, $\mathbf{a}_\mathrm{out}( \omega ) = \mathbf{\Xi}( \omega ) \mathbf{a}_\mathrm{in}( \omega )$, calculated from linearized equations of motion following Ref.~\cite{andrews_bidirectional_2014}.
Here
\begin{equation}
\mathbf{a}_\mathrm{in} = \left( \hat{ a }_\mathrm{in,B}, \hat{ a }_\mathrm{in,F}, \hat{ a }_\mathrm{in,int}, \hat{ b }_\mathrm{in,ex}, \hat{ b }_\mathrm{in,int}, \hat{c}_\mathrm{in}, 
\hat{ a }^\dagger_\mathrm{in,B}, \hat{ a }^\dagger_\mathrm{in,F}, \hat{ a }^\dagger_\mathrm{in,int}, \hat{ b }^\dagger_\mathrm{in,ex}, \hat{ b }^\dagger_\mathrm{in,int}, \hat{c}^\dagger_\mathrm{in} \right)^T
\end{equation}
is a vector of input fields for optical ($\hat{a}_\mathrm{in}$), microwave ($\hat{b}_\mathrm{in}$), and mechanical ($\hat{c}_\mathrm{in}$) modes.

In terms of input fields, the spectral matrix is then
\begin{equation} \label{eq:sinfields}
\mathbf{C}( \omega ) \delta( \omega - \omega' ) = \frac{1}{2} \mathbf{\Xi}^*( \omega' )  \left(  \langle [ \mathbf{a}_\mathrm{in} (\omega') ]^\dagger \mathbf{a}^T_\mathrm{in}( \omega ) \rangle + 
\langle [ \mathbf{a}_\mathrm{in}( \omega ) [ \mathbf{a}^T_\mathrm{in} (\omega') ]^{\dagger} ]^T \rangle \right) \mathbf{\Xi}^T( \omega ),
\end{equation}
where transposes act only on matrices and adjoints act only on operators.
The input fields are thermal states, satisfying for example $\langle [ c_\mathrm{in}( \omega' ) ]^\dagger c_\mathrm{in}( \omega ) \rangle = \delta( \omega - \omega' ) n_\mathrm{th,m}$ and $\langle c_\mathrm{in}( \omega ) [ c_\mathrm{in}( \omega' ) ]^\dagger \rangle = \delta( \omega - \omega' ) ( n_\mathrm{th,m} + 1 )$, where $n_\mathrm{th,m}$ is the number of thermal phonons in the mechanical oscillator.
Integrating over $\omega'$, Eq.~\ref{eq:sinfields} can then be rewritten
\begin{equation} \label{eq:cfinal}
\mathbf{C}( \omega ) = \frac{1}{2} \mathbf{\Xi}^*( \omega ) \mathbf{\Xi}^T( \omega ) +  \mathbf{\Xi}^*( \omega ) \mathbf{\Sigma} \mathbf{\Xi}^T( \omega ).
\end{equation}
The first term originates from quantum noise of the input modes, whereas the second is a thermal contribution that vanishes at zero temperature, 
due to the fact that
\begin{equation}
\mathbf{\Sigma} = 
\begin{pmatrix}
\mathbf{N} & 0 \\
0 & \mathbf{N}
\end{pmatrix},
\end{equation}
where $\mathbf{N} = \mathrm{Diag}[ n_\mathrm{th,o}, n_\mathrm{th,o}, n_\mathrm{th,o}, n_\mathrm{th,e}, n_\mathrm{th,e}, n_\mathrm{th,m}]$ is a diagonal matrix of thermal noise contributed by each input mode.

Note that the diagonal entries of $\mathbf{C}( \omega )$ are real, whereas the off-diagonal components are in general complex. 
In terms of quantities measured in the main text, for example, $[ \hat{ a }_\mathrm{out,F}( \omega ) ]^\dagger \hat{ a }_\mathrm{out,F}( \omega ) = X_\mathrm{o}^2 + Y_\mathrm{o}^2$, whereas
\begin{equation}
[ \hat{ a }_\mathrm{out,F}( \omega ) ]^\dagger \hat{ b }_\mathrm{out}( \omega )  = X_\mathrm{o} X_\mathrm{e} + Y_\mathrm{o} Y_\mathrm{e} + i X_\mathrm{o} Y_\mathrm{e} - i Y_\mathrm{o} X_\mathrm{e}.
\end{equation}
In practice one can adjust the optical and microwave demodulator phases to make all entries real, which corresponds to removing $X$-$Y$ microwave-optical correlations, $\langle X_\mathrm{o} Y_\mathrm{e} \rangle = \langle Y_\mathrm{o} X_\mathrm{e} \rangle = 0$.
In the experiment, the demodulator phases are adjusted to approximately fulfill this condition.


While the full expressions for the matrix elements of Eq.~\ref{eq:cfinal} are cumbersome, in the limit of optimal detuning ($\Delta_\mathrm{o} = \Delta_\mathrm{e} = -\omega_\mathrm{m}$), weak damping ($\Gamma_\mathrm{e},\Gamma_\mathrm{o} \ll \kappa_\mathrm{e},\kappa_\mathrm{o}$), and resolved sidebands ($\kappa_\mathrm{o}, \kappa_\mathrm{e} \ll \omega_\mathrm{m}$), on mechanical resonance the electro-optic correlations take the simple form
\begin{eqnarray} 
\mathbf{C}_\mathrm{eo}( \omega_\mathrm{m} ) &=& \frac{1}{2} \mathbb{I}_{2} + \mathbf{C}^{(\mathrm{th})}( \omega_\mathrm{m} ), \\
\mathbf{C}^{(\mathrm{th})}( \omega_\mathrm{m} ) &=& \frac{4 n_\mathrm{th,m} \gamma_\mathrm{m} }{ ( \Gamma_\mathrm{o} + \Gamma_\mathrm{e} + \gamma_\mathrm{m} )^2 }
\begin{pmatrix}
\Gamma_\mathrm{o} & \sqrt{\Gamma_\mathrm{o} \Gamma_\mathrm{e}} \\
\sqrt{\Gamma_\mathrm{e} \Gamma_\mathrm{o}} & \Gamma_\mathrm{e}
\end{pmatrix}, \label{eq:ctherm}
\end{eqnarray}
where $\mathbb{I}_2$ is the 2x2 identity, representing the contribution of quantum noise.
The diagonal elements of $\mathbf{C}^{\mathrm{(th)}}$ have a straightforward interpretation. 
Thermal noise added to the optical output mode, $C^{(\mathrm{th})}_{1,1}$, is due to the resonant transmission of mechanical noise $n_\mathrm{th,m}$ from the bath, coupled at rate $\gamma_\mathrm{m}$, to the optical mode, coupled at rate $\Gamma_\mathrm{o}$.
Similarly, thermal noise added to the microwave output mode, $C^{(\mathrm{th})}_{2,2}$, is due to the resonant transmission of mechanical noise from the bath to the microwave mode, coupled at rate $\Gamma_\mathrm{e}$.
Since the mechanical resonator couples to the optical, microwave, and bath modes, its total linewidth is $\Gamma_\mathrm{o} + \Gamma_\mathrm{e} + \gamma_\mathrm{m}$.

The elements of $\mathbf{C}^{(\mathrm{th})}$ satisfy the property
\begin{equation} \label{eq:maxcorr}
C^{(\mathrm{th})}_{1,2} = \sqrt{ C^{(\mathrm{th})}_{1,1} C^{(\mathrm{th})}_{2,2} }.
\end{equation}
As a consequence, the smallest eigenvalue of $\mathbf{C}_\mathrm{eo}$ is equal to $1/2~\mathrm{photon}$.
If $C^{(\mathrm{th})}_{1,2}$ were to exceed $\sqrt{ C^{(\mathrm{th})}_{1,1} C^{(\mathrm{th})}_{2,2} }$ in magnitude, then the smallest eigenvalue of $\mathbf{C}_\mathrm{eo}$ would be less than $1/2~\mathrm{photon}$, signaling the presence of two-mode squeezing.
Equation~\ref{eq:maxcorr} therefore represents the condition for maximal classical correlations, as they are as large as possible without implying the presence of squeezed quantum noise.

While Eq.~\ref{eq:ctherm} is derived under restricted assumptions, we have verified numerically that Eq.~\ref{eq:maxcorr} is satisfied for arbitrary detunings and linewidths, including for the system parameters reported in Table~\ref{tab:params}.
Interestingly, the property $C^{(\mathrm{th})}_{1,1} = C^{(\mathrm{th})}_{2,2}$ can also be obtained under more general conditions, provided an appropriate detuning is used.
For instance, if we had instead chosen $\Delta_\mathrm{o} = -0.38~\mathrm{MHz}$ but kept all other parameters the same as Table~\ref{tab:params} we would have obtained $C^{(\mathrm{th})}_{1,1} = C^{(\mathrm{th})}_{2,2}$ at the expense of weaker optomechanical damping per incident photon.

Importantly, imperfect sideband resolution does not weaken the strength of the classical correlations.
With imperfect sideband resolution, arbitrary detunings, and cavity loss, the correlation matrix still can still be written in the form
\begin{equation}
\mathbf{C}_\mathrm{eo}( \omega_\mathrm{m} ) = \frac{1}{2} \mathbb{I}_{2} + \mathbf{C}^{'}( \omega_\mathrm{m} ),
\end{equation}
with $\mathbf{C}^{'}$ satisfying
\begin{equation} \label{eq:maxcorrSB}
C^{\mathrm{'}}_{1,2} = \sqrt{ C^{'}_{1,1} C^{'}_{2,2} }.
\end{equation}
The correlations are still classically maximal, but, in contrast to Eq.~(\ref{eq:maxcorr}), there are now contributions both from thermal and quantum noise.

\section{Feedforward theory}

After passing through a sideband-resolved and impedance-matched converter, the signal and ancilla states are described by the quadratures 
\begin{eqnarray} \label{eq:quadDef1}
q_\mathrm{s,out} = \sqrt{ \eta } q_\mathrm{s,in} + \sqrt{ 1 - \eta } V_1 + q_\mathrm{th,out},  \\
q_\mathrm{a,out} = \sqrt{ \eta } q_\mathrm{a,in} + \sqrt{ 1 - \eta } V_2 + q_\mathrm{th,out}, \label{eq:quadDef2}
\end{eqnarray}
where $q_\mathrm{s,in}$ and $q_\mathrm{a,in}$ are the input signal and ancilla quadratures, $V_1$ and $V_2$ are vacuum noise introduced by imperfect conversion, and $q_\mathrm{th,out}$ is the thermal noise in the output spectrum. 
It has been assumed that the pumps are configured to give identical thermal noise on microwave and  optical  outputs. 
For vacuum signal and ancilla, the quadratures in Eqs.~(\ref{eq:quadDef1}-\ref{eq:quadDef2}) give a correlation matrix of the appropriate form, as discussed in Sec.~\ref{sec:corrtheory}.
Assuming noiseless single-quadrature measurement and perfect converter efficiency, one can construct the fed-forward quadrature $\check{ q }_\mathrm{s} = q_\mathrm{s} - w q_\mathrm{a}$. 
Choosing $w = 1$ yields
\begin{equation}
\check{ q }_\mathrm{s} = q_\mathrm{s,q} - q_\mathrm{a,q},
\end{equation}
so thermal noise can be completely eliminated using feedforward.
With a non-ideal measurement or converter, some additional noise will be fed forward along with the quantum noise of the ancilla; $w=1$ will not in general be the optimal value for cancelling the thermal noise. 

In the more realistic case of imperfect sideband resolution and imperfect cavities, gain and loss terms are introduced which cause an impedance mismatch.
Due to the mismatch, some signal is reflected at peak efficiency.
Reflected signals are inevitably fed forward along with the correlated noise, and feedforward is therefore degraded.
Reflections can be reduced by operating away from peak efficiency, but this also degrades feedforward.
Interestingly, an adaptive control strategy that avoids the matching requirement has recently been proposed \cite{zhang_quantum_2018}.
The compatibility of this technique with feedforward, and other considerations for optimally handling imperfect sideband resolution, are interesting problems for future investigation.

Signal reflections are small for the feedforward experiment described in the main text, and feedforward performance is overwhelmingly limited by measurement efficiency.
Equations (\ref{eq:quadDef1}-\ref{eq:quadDef2}) therefore represent a useful limit for understanding the experiment and its main limitations. A full treatment should include explicitly the effects of gain and loss, and thoroughly explore the tradeoffs for optimizing feedforward, which remains an open problem. It should be emphasized that the experimental characterization of feedforward performance in the main text does not require any assumptions about sideband resolution: we explicitly measure the variance of the fed-forward optical quadrature and observe that it is reduced.

\subsection{Gaussian states}

For a Gaussian signal and ancilla, one need only keep track of quadrature variances.
For the ideal case introduced above, the variance of the fed-forward quadrature is
\begin{equation}
\langle \check{ q }_\mathrm{s}^2 \rangle = \langle q_\mathrm{s,q}^2 \rangle + \langle q_\mathrm{a,q}^2 \rangle.
\end{equation}
When the ancilla is simply vacuum, feedforward adds vacuum noise, while removing all thermal noise.
This is the classical feedforward discussed in the main text.
If a squeezed ancilla is chosen, then in the limit of perfect squeezing where $\langle q_\mathrm{a,q}^2 \rangle=0$, the squeezed quadrature can be fed forward noiselessly.
This enables upconversion of a squeezed state or noiseless measurement of a remotely prepared microwave quadrature.

The squeezed protocol requires a threshold converter efficiency to be reached.
For a finite conversion efficiency, but with an otherwise perfect homodyne measurement apparatus, the fed-forward variance is
\begin{equation}
\langle \check{ q }_\mathrm{s}^2 \rangle = \eta \langle q_\mathrm{s,in}^2 \rangle + \eta \langle q_\mathrm{a,in}^2 \rangle + 2 ( 1 - \eta ) \langle V^2 \rangle,
\end{equation}
where $q_\mathrm{s,in}$ is the input signal quadrature and $q_\mathrm{a,in}$ is the input ancilla quadrature, and $\langle V^2 \rangle$ denotes the vacuum variance.
In the limit of perfect ancilla squeezing, $\langle q_\mathrm{a,in}^2 \rangle = 0$, less than one vacuum of noise is only added when $\eta > 1/2$.
This is a threshold for measuring squeezing in an upconverted signal, or more generally measuring a remotely prepared microwave state with added noise less than $1/2~\mathrm{photon}$, referred to as a quantum threshold in the main text.

If one considers separate efficiency thresholds for microwave-optical and optical-microwave conversion and operates in a regime where thermal noise is added asymmetrically, it is possible to relax the threshold $\eta=1/2$.
If much less noise is added to one output port than the other, performance is improved when the low-noise output is used for the signal, so a lower efficiency can be tolerated in that direction.
However, if the high-noise output is used for the signal, a higher efficiency is required.
Strictly, $\eta=1/2$ is therefore the minimum threshold for bi-directional conversion of squeezed states.


\subsection{Qubit up-conversion}
With a squeezed ancilla, the feedforward protocol can be used to remove noise in one quadrature.
The task then is to encode a qubit in a single quadrature, and to find an encoding that makes this qubit robust to noise in that same quadrature. 
Here we examine one encoding to demonstrate that upconversion of a quantum signal in the low-cooperativity limit is possible in principle, without regard to technical feasibility. 
While our encoding has the virtue of formal simplicity, it is unlikely to be optimal with regard to efficiency requirements. 
We speculate that a Gottesman-Kitaev-Preskill qubit encoding may improve efficiency thresholds for qubit conversion \cite{gottesman_encoding_2001,albert_performance_2017}.


Consider using a squeezed ancilla to eliminate classical noise in the $Y$ quadrature, leaving classical noise in the $X$ quadrature. 
The noise operator $\cN$ is of the form
\begin{align} \label{Eq:Xshift}
\cN(\rho) = \int  U_\beta \rho U_\beta^\dagger p(\beta){\rm d}\beta,
\end{align}
where $\rho$ is the density operator of the qubit, $U_\beta|x\rangle \rightarrow |x+\beta\rangle$, and
\begin{align} 
p(\beta) = \frac{1}{\sqrt{2\pi}\sigma_X}e^{-\beta^2/(2\sigma_X^2)}.
\end{align}

Our goal is to show that, with a suitable decoder, this channel can transmit a qubit.
This is equivalent to showing the channel can transmit a half of a maximally entangled state with high fidelity.
It is formally simplest to demonstrate the required fidelity for an encoding expressed in terms of $X$ eigenstates, $X|x\rangle = x |x\rangle$, but similar encodings for coherent states should work as well.

A decoder should preserve the quantum state, but should allow for information to be gained about $\beta$ so that the state can be shifted back into its original space.
What we want now is a pair of logical states $|\bar{0}\rangle$ and $|\bar{1}\rangle$ and a decoding operation $\cD: H \rightarrow A$, where $H$ is the output space of $\cN$ and $A$ is a single-qubit space (spanned by $\ket{\bar{0}}$ and $\ket{\bar{1}}$), such that the entanglement fidelity
\begin{align}
F_\mathrm{ent} = \bra{\phi_0} I \otimes (\cD\circ\cN)(\ket{\phi_0}\bra{\phi_0})\ket{\phi_0} \approx 1,
\end{align}
where $\ket{\phi_0} = \frac{1}{\sqrt{2}}(\ket{00} + \ket{11})$.  
This fidelity can be rewritten as

\begin{align}\label{Eq:Fidelity}
F_\mathrm{ent} = \frac{1}{4} \sum_{s=0}^1\sum_{t=0}^1 \bra{\bar{s}}\cD\circ\cN(\ket{\bar{s}}\bra{\bar{t}})\ket{\bar{t}},
\end{align}
which will make it easier to calculate.

The encoding we will choose depends on an $X$ eigenvalue $b>0$, and is of the form
\begin{align}
\ket{\bar{0}}& = \ket{b}\\
\ket{\bar{1}}& = \ket{-b}.
\end{align}
Noise of the form Eq.~(\ref{Eq:Xshift}) will shift this codespace to a new space spanned by $\ket{b+\beta}$ and $\ket{-b+\beta}$.  Choosing $b\gg\sigma_X$ will enable us to figure out the displacement $\beta$ and shift the codespace back.

An excellent observable to measure to determine the value of $\beta$ is
\begin{align}
A_b = \int_{-b}^b \gamma P_\gamma \rm{d}\gamma.
\end{align}
Here, $P_\gamma = |b+\gamma\rangle\langle b+\gamma| +|-b+\gamma\rangle\langle -b+\gamma|$ is the projector onto the codespace after being displaced by $\gamma$
in the $x$ direction.  Given outcome $\gamma$ (which we will show below will almost surely equal the true shift $\beta$), our decoding will be completed
by applying a conditional displacement $U_\gamma$ to map $P_\gamma$ back to the original codespace $P_0$.

\begin{figure}
\begin{minipage}{\linewidth}
\scalebox{1}{\includegraphics{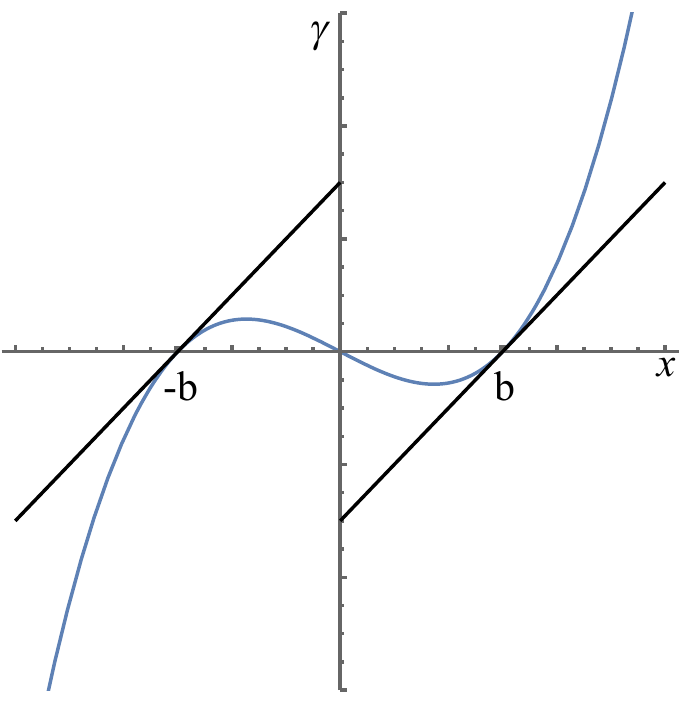}}
\caption{Plot of measurement outcome for ideal observable $f_b(x)$ (black) and approximated observable $\tilde{f}_b(x)$ (blue).}
\label{cubic}
\end{minipage}
\end{figure}

It is useful to re-express $A_b$ in terms of the $X$ operator as follows:

\begin{align}
A_b & = \int_{-b}^b \gamma P_\gamma \rm{d}\gamma\\
& = \int_{-b}^b \gamma\left( \proj{b+\gamma}+ \proj{-b+\gamma}\right)\rm{d}\gamma\\
& = \int_{-b}^b \gamma \proj{b+\gamma}\rm{d}\gamma + \int_{-b}^b \gamma \proj{-b+\gamma}\rm{d}\gamma\\
& = \int_{0}^{2b} (x-b) \proj{x}\rm{d}x +\int_{-2b}^{0} (x+b) \proj{x}\rm{d}x \\
& = \int_{-2b}^{2b}f_b(x)\proj{x}\rm{d}x\\
& = \int_{-\infty}^{\infty}f_b(x)\proj{x}\rm{d}x\\
& = f_b(X),
\end{align}
where
\begin{align}
        f_b(x) & = \begin{cases}
                        x-b \text{   for } 2b \geq x\geq 0 \\
                        x+b \text{   for } 0 \geq x \geq -2b\\
                        0   \text{   for } |x|> 2b.
                    \end{cases}
\end{align}
The function $f_b(x)$ is plotted in Fig.~\ref{cubic}.

To derive a result for the entanglement fidelity we would like to evaluate Eq.~(\ref{Eq:Fidelity}). The following calculation will be useful: fixing $\ket{\bar{s}}$
and requiring $|\beta|<b$ and $|\gamma|<b$, we have
\begin{align}
\bra{\bar{s}}U_\gamma^\dagger P_\gamma U_\beta \ket{\bar{s}} & = \bra{\bar{s}} \left(\ket{b}\bra{b+\gamma}+ \ket{-b}\bra{-b+\gamma}\right)U_\beta\ket{\bar{s}}\\
& = \bra{(-1)^s b} \left(\ket{b}\bra{b+\gamma}+ \ket{-b}\bra{-b+\gamma}\right)\ket{\beta + (-1)^{s}b}\\
& = \braket{(-1)^s b + \gamma}{\beta+(-1)^sb}\\
& = \braket{\gamma}{\beta}.
\end{align}

Our decoder $\cD$ is implemented by a two-step process.  The first step is the 
quantum nondemolition measurement of $A_b = f_{b}(X)$ which results in an estimate $\gamma$ of the noise $\beta$ that was applied.  The second step is a translation $U_\gamma^\dagger$ to map back to the original codespace.  We now evaluate the fidelity achieved by this decoding.  To evaluate Eq.(\ref{Eq:Fidelity}), we first fix $\ket{\bar{s}}$ and $\ket{\bar{t}}$,and compute
\begin{align}
\bra{\bar{s}}\cD\circ\cN(\ket{\bar{s}}\bra{\bar{t}})\ket{\bar{t}} & 
\geq \int_{|\beta|<b}\int_{|\gamma|<b}\bra{\bar{s}}U_\gamma^\dagger P_\gamma U_\beta \ket{\bar{s}}\bra{\bar{t}} U_\beta^\dagger P_\gamma U_\gamma\ket{\bar{t}} p(\beta) \rm{d}\gamma \rm{d}\beta\\
& = \int_{|\beta|<b}\int_{|\gamma|<b}  \braket{\gamma}{\beta}   \braket{\beta}{\gamma} p(\beta) \rm{d}\gamma \rm{d}\beta\\
& =  \int_{|\beta|<b}p(\beta)\rm{d}\beta\\
& =  \int_{|\beta|<b}\frac{1}{\sqrt{2\pi}}\exp\left(-\frac{\beta^2}{2\sigma_X^2}\right)\rm{d}\beta\\
& = 1 - \textbf{erfc}\left(\frac{b}{\sigma_X}\right)\\
& = 1-O\left(\exp\left(-\frac{\beta^2}{2\sigma_X^2}\right)\right).
\end{align}

Averaging over $\bar{s}$ and $\bar{t}$ gives us the result

\begin{align}
F_\mathrm{ent} & \geq 1- O(\exp(-b^2/(2\sigma_X^2))),
\end{align}
which will be close to 1, for $b\gg\sigma_X$. In other words, if the displacement $b$ far exceeds thermal noise, errors can be diagnosed and corrected with an appropriately chosen decoding operation.

A key step in the decoding procedure described above is the QND measurement of $f_b(X)$, but implementing this measurement could be a challenge.  
We therefore consider a more physical decoding procedure, and argue that it will give similar fidelity to the one described above.  In particular, we let 
\begin{align}
\tilde{f}_b(x) = \frac{1}{2b^2}x^3-\frac{1}{2}x.
\end{align}
This is a good approximation to $f_b$ near $\pm b$ (see Fig.~\ref{cubic}): for $|\gamma|\ll b$  we have $\tilde{f}_b(\pm b+ \gamma) = \gamma$, which is exactly what we require.  Because of our choice $b\gg \sigma_X$, we are very likely to get a measurement outcome in this range if we measure $\tilde{f}_b(X)$, so doing so is a good proxy for measuring $f_b(X)$.  

Our decoding procedure is again two-step:  First, do a QND measurement of 
\begin{align}
    \tilde{A}_b = \frac{1}{2b^2}X^3-\frac{1}{2}X,
\end{align}
obtaining a result $\gamma$.  Second, apply $U_\gamma^\dagger$.  This gives a transmitted fidelity close to 1 as long as $b\gg \sigma_X$.

\begin{thebibliography}{10}
\expandafter\ifx\csname url\endcsname\relax
  \def\url#1{\texttt{#1}}\fi
\expandafter\ifx\csname urlprefix\endcsname\relax\def\urlprefix{URL }\fi
\providecommand{\bibinfo}[2]{#2}
\providecommand{\eprint}[2][]{\url{#2}}

\bibitem{cirac_quantum_1997}
\bibinfo{author}{Cirac, J.~I.}, \bibinfo{author}{Zoller, P.},
  \bibinfo{author}{Kimble, H.~J.} \& \bibinfo{author}{Mabuchi, H.}
\newblock \bibinfo{title}{Quantum state transfer and entanglement distribution
  among distant nodes in a quantum network}.
\newblock \emph{\bibinfo{journal}{Phys. Rev. Lett.}}
  \textbf{\bibinfo{volume}{78}}, \bibinfo{pages}{3221--3224}
  (\bibinfo{year}{1997}).

\bibitem{ekert_quantum_1991}
\bibinfo{author}{Ekert, A.~K.}
\newblock \bibinfo{title}{{Quantum cryptography based on Bell's theorem}}.
\newblock \emph{\bibinfo{journal}{Phys. Rev. Lett.}}
  \textbf{\bibinfo{volume}{67}}, \bibinfo{pages}{661--663}
  (\bibinfo{year}{1991}).

\bibitem{cirac_distributed_1999}
\bibinfo{author}{Cirac, J.~I.}, \bibinfo{author}{Ekert, A.~K.},
  \bibinfo{author}{Huelga, S.~F.} \& \bibinfo{author}{Macchiavello, C.}
\newblock \bibinfo{title}{{Distributed quantum computation over noisy
  channels}}.
\newblock \emph{\bibinfo{journal}{Phys. Rev. A}} \textbf{\bibinfo{volume}{59}},
  \bibinfo{pages}{4249--4254} (\bibinfo{year}{1999}).

\bibitem{Dur_entanglement_2003}
\bibinfo{author}{D\"ur, W.} \& \bibinfo{author}{Briegel, H.-J.}
\newblock \bibinfo{title}{Entanglement purification for quantum computation}.
\newblock \emph{\bibinfo{journal}{Phys. Rev. Lett.}}
  \textbf{\bibinfo{volume}{90}}, \bibinfo{pages}{067901}
  (\bibinfo{year}{2003}).

\bibitem{kelly_state_2015}
\bibinfo{author}{Kelly, J.}, \bibinfo{author}{Barends, R.},
  \bibinfo{author}{Fowler, A.~G.}, \bibinfo{author}{Megrant, A.} \&
  \bibinfo{author}{Jeffrey, E.}
\newblock \bibinfo{title}{{State preservation by repetitive error detection in
  a superconducting quantum circuit}}.
\newblock \emph{\bibinfo{journal}{Nature}} \textbf{\bibinfo{volume}{519}},
  \bibinfo{pages}{66--69} (\bibinfo{year}{2015}).

\bibitem{ofek_extending_2016}
\bibinfo{author}{Ofek, N.} \emph{et~al.}
\newblock \bibinfo{title}{{Extending the lifetime of a quantum bit with error
  correction in superconducting circuits}}.
\newblock \emph{\bibinfo{journal}{Nature}} \textbf{\bibinfo{volume}{536}},
  \bibinfo{pages}{441--445} (\bibinfo{year}{2016}).

\bibitem{verdu_strong_2009}
\bibinfo{author}{Verd{\'u}, J.} \emph{et~al.}
\newblock \bibinfo{title}{Strong magnetic coupling of an ultracold gas to a
  superconducting waveguide cavity}.
\newblock \emph{\bibinfo{journal}{Phys. Rev. Lett.}}
  \textbf{\bibinfo{volume}{103}}, \bibinfo{pages}{043603}
  (\bibinfo{year}{2009}).

\bibitem{hafezi_atomic_2012}
\bibinfo{author}{Hafezi, M.} \emph{et~al.}
\newblock \bibinfo{title}{Atomic interface between microwave and optical
  photons}.
\newblock \emph{\bibinfo{journal}{Phys. Rev. A}} \textbf{\bibinfo{volume}{85}},
  \bibinfo{pages}{020302} (\bibinfo{year}{2012}).

\bibitem{imamoglu_cavity_2009}
\bibinfo{author}{Imamoglu, A.}
\newblock \bibinfo{title}{Cavity {QED} based on collective magnetic dipole
  coupling: Spin ensembles as hybrid two-level systems}.
\newblock \emph{\bibinfo{journal}{Phys. Rev. Lett.}}
  \textbf{\bibinfo{volume}{102}}, \bibinfo{pages}{083602}
  (\bibinfo{year}{2009}).

\bibitem{marcos_coupling_2010}
\bibinfo{author}{Marcos, D.} \emph{et~al.}
\newblock \bibinfo{title}{Coupling nitrogen-vacancy centers in diamond to
  superconducting flux qubits}.
\newblock \emph{\bibinfo{journal}{Phys. Rev. Lett.}}
  \textbf{\bibinfo{volume}{105}}, \bibinfo{pages}{210501}
  (\bibinfo{year}{2010}).

\bibitem{kubo_strong_2010}
\bibinfo{author}{Kubo, Y.} \emph{et~al.}
\newblock \bibinfo{title}{Strong coupling of a spin ensemble to a
  superconducting resonator}.
\newblock \emph{\bibinfo{journal}{Phys. Rev. Lett.}}
  \textbf{\bibinfo{volume}{105}}, \bibinfo{pages}{140502}
  (\bibinfo{year}{2010}).

\bibitem{williamson_magneto-optic_2014}
\bibinfo{author}{Williamson, L.~A.}, \bibinfo{author}{Chen, Y.-H.} \&
  \bibinfo{author}{Longdell, J.~J.}
\newblock \bibinfo{title}{Magneto-optic modulator with unit quantum
  efficiency}.
\newblock \emph{\bibinfo{journal}{Phys. Rev. Lett.}}
  \textbf{\bibinfo{volume}{113}}, \bibinfo{pages}{203601}
  (\bibinfo{year}{2014}).

\bibitem{hisatomi_bidirectional_2016}
\bibinfo{author}{Hisatomi, R.} \emph{et~al.}
\newblock \bibinfo{title}{{Bidirectional conversion between microwave and light
  via ferromagnetic magnons}}.
\newblock \emph{\bibinfo{journal}{Phys. Rev. B}} \textbf{\bibinfo{volume}{93}},
  \bibinfo{pages}{174427} (\bibinfo{year}{2016}).

\bibitem{ilchenko_whispering-gallery-mode_2003}
\bibinfo{author}{Ilchenko, V.~S.}, \bibinfo{author}{Savchenkov, A.~A.},
  \bibinfo{author}{Matsko, A.~B.} \& \bibinfo{author}{Maleki, L.}
\newblock \bibinfo{title}{Whispering-gallery-mode electro-optic modulator and
  photonic microwave receiver}.
\newblock \emph{\bibinfo{journal}{J. Opt. Soc. Am. B}}
  \textbf{\bibinfo{volume}{20}}, \bibinfo{pages}{333--342}
  (\bibinfo{year}{2003}).

\bibitem{xiong_low-loss_2012}
\bibinfo{author}{Xiong, C.}, \bibinfo{author}{Pernice, W. H.~P.} \&
  \bibinfo{author}{Tang, H.~X.}
\newblock \bibinfo{title}{Low-loss, silicon integrated, aluminum nitride
  photonic circuits and their use for electro-optic signal processing}.
\newblock \emph{\bibinfo{journal}{Nano Lett.}} \textbf{\bibinfo{volume}{12}},
  \bibinfo{pages}{3562--3568} (\bibinfo{year}{2012}).

\bibitem{rueda_efficient_2016}
\bibinfo{author}{Rueda, A.} \emph{et~al.}
\newblock \bibinfo{title}{Efficient microwave to optical photon conversion: an
  electro-optical realization}.
\newblock \emph{\bibinfo{journal}{Optica}} \textbf{\bibinfo{volume}{3}},
  \bibinfo{pages}{597--604} (\bibinfo{year}{2016}).

\bibitem{bochmann_nanomechanical_2013}
\bibinfo{author}{Bochmann, J.}, \bibinfo{author}{Vainsencher, A.},
  \bibinfo{author}{Awschalom, D.~D.} \& \bibinfo{author}{Cleland, A.~N.}
\newblock \bibinfo{title}{Nanomechanical coupling between microwave and optical
  photons}.
\newblock \emph{\bibinfo{journal}{Nat. Phys.}} \textbf{\bibinfo{volume}{9}},
  \bibinfo{pages}{712--716} (\bibinfo{year}{2013}).

\bibitem{andrews_bidirectional_2014}
\bibinfo{author}{Andrews, R.~W.} \emph{et~al.}
\newblock \bibinfo{title}{Bidirectional and efficient conversion between
  microwave and optical light}.
\newblock \emph{\bibinfo{journal}{Nat. Phys.}} \textbf{\bibinfo{volume}{10}},
  \bibinfo{pages}{321--326} (\bibinfo{year}{2014}).

\bibitem{vainsencher_bi-directional_2016}
\bibinfo{author}{Vainsencher, A.}, \bibinfo{author}{Satzinger, K.~J.},
  \bibinfo{author}{Peairs, G.~A.} \& \bibinfo{author}{Cleland, A.~N.}
\newblock \bibinfo{title}{{Bi-directional conversion between microwave and
  optical frequencies in a piezoelectric optomechanical device}}.
\newblock \emph{\bibinfo{journal}{Appl. Phys. Lett.}}
  \textbf{\bibinfo{volume}{109}}, \bibinfo{pages}{033107}
  (\bibinfo{year}{2016}).

\bibitem{bagci_optical_2014}
\bibinfo{author}{Bagci, T.} \emph{et~al.}
\newblock \bibinfo{title}{Optical detection of radio waves through a
  nanomechanical transducer}.
\newblock \emph{\bibinfo{journal}{Nature}} \textbf{\bibinfo{volume}{507}},
  \bibinfo{pages}{81--85} (\bibinfo{year}{2014}).

\bibitem{takeda_electro-mechano-optical_2018}
\bibinfo{author}{Takeda, K.} \emph{et~al.}
\newblock \bibinfo{title}{Electro-mechano-optical detection of nuclear magnetic
  resonance}.
\newblock \emph{\bibinfo{journal}{Optica}} \textbf{\bibinfo{volume}{5}},
  \bibinfo{pages}{152--158} (\bibinfo{year}{2018}).

\bibitem{safavi-naeini_proposal_2011}
\bibinfo{author}{Safavi-Naeini, A.~H.} \& \bibinfo{author}{Painter, O.}
\newblock \bibinfo{title}{Proposal for an optomechanical traveling wave
  phonon--photon translator}.
\newblock \emph{\bibinfo{journal}{New J. Phys.}} \textbf{\bibinfo{volume}{13}},
  \bibinfo{pages}{013017} (\bibinfo{year}{2011}).

\bibitem{hill_coherent_2012}
\bibinfo{author}{Hill, J.~T.}, \bibinfo{author}{Safavi-Naeini, A.~H.},
  \bibinfo{author}{Chan, J.} \& \bibinfo{author}{Painter, O.}
\newblock \bibinfo{title}{Coherent optical wavelength conversion via cavity
  optomechanics}.
\newblock \emph{\bibinfo{journal}{Nat. Commun.}} \textbf{\bibinfo{volume}{3}},
  \bibinfo{pages}{1196} (\bibinfo{year}{2012}).

\bibitem{zeuthen_figures_2016}
\bibinfo{author}{Zeuthen, E.}, \bibinfo{author}{Schliesser, A.},
  \bibinfo{author}{S{\o}rensen, A.~S.} \& \bibinfo{author}{Taylor, J.~M.}
\newblock \bibinfo{title}{{Figures of merit for quantum transducers}}.
\newblock \eprint{Preprint at http://arxiv.org/abs/1610.01099 (2016).}

\bibitem{safavi-naeini_design_2010}
\bibinfo{author}{Safavi-Naeini, A.~H.} \& \bibinfo{author}{Painter, O.}
\newblock \bibinfo{title}{Design of optomechanical cavities and waveguides on a
  simultaneous bandgap phononic-photonic crystal slab}.
\newblock \emph{\bibinfo{journal}{Opt. Express}} \textbf{\bibinfo{volume}{18}},
  \bibinfo{pages}{14926--14943} (\bibinfo{year}{2010}).

\bibitem{cohadon_cooling_1999}
\bibinfo{author}{Cohadon, P.~F.}, \bibinfo{author}{Heidmann, A.} \&
  \bibinfo{author}{Pinard, M.}
\newblock \bibinfo{title}{Cooling of a mirror by radiation pressure}.
\newblock \emph{\bibinfo{journal}{Phys. Rev. Lett.}}
  \textbf{\bibinfo{volume}{83}}, \bibinfo{pages}{3174--3177}
  (\bibinfo{year}{1999}).

\bibitem{arcizet_high-sensitivity_2006}
\bibinfo{author}{Arcizet, O.} \emph{et~al.}
\newblock \bibinfo{title}{High-sensitivity optical monitoring of a
  micromechanical resonator with a quantum-limited optomechanical sensor}.
\newblock \emph{\bibinfo{journal}{Phys. Rev. Lett.}}
  \textbf{\bibinfo{volume}{97}}, \bibinfo{pages}{133601}
  (\bibinfo{year}{2006}).

\bibitem{poggio_feedback_2007}
\bibinfo{author}{Poggio, M.}, \bibinfo{author}{Degen, C.~L.},
  \bibinfo{author}{Mamin, H.~J.} \& \bibinfo{author}{Rugar, D.}
\newblock \bibinfo{title}{Feedback cooling of a cantilever's fundamental mode
  below 5~mk}.
\newblock \emph{\bibinfo{journal}{Phys. Rev. Lett.}}
  \textbf{\bibinfo{volume}{99}}, \bibinfo{pages}{017201}
  (\bibinfo{year}{2007}).

\bibitem{genes_ground-state_2008}
\bibinfo{author}{Genes, C.}, \bibinfo{author}{Vitali, D.},
  \bibinfo{author}{Tombesi, P.}, \bibinfo{author}{Gigan, S.} \&
  \bibinfo{author}{Aspelmeyer, M.}
\newblock \bibinfo{title}{Ground-state cooling of a micromechanical oscillator:
  {Comparing} cold damping and cavity-assisted cooling schemes}.
\newblock \emph{\bibinfo{journal}{Phys. Rev. A}} \textbf{\bibinfo{volume}{77}},
  \bibinfo{pages}{033804} (\bibinfo{year}{2008}).

\bibitem{wilson_measurement-based_2015}
\bibinfo{author}{Wilson, D.~J.} \emph{et~al.}
\newblock \bibinfo{title}{{Measurement-based control of a mechanical oscillator
  at its thermal decoherence rate}}.
\newblock \emph{\bibinfo{journal}{Nature}} \textbf{\bibinfo{volume}{524}},
  \bibinfo{pages}{325--329} (\bibinfo{year}{2015}).

\bibitem{rossi_enhancing_2017}
\bibinfo{author}{Rossi, M.} \emph{et~al.}
\newblock \bibinfo{title}{Enhancing sideband cooling by feedback-controlled
  light}.
\newblock \emph{\bibinfo{journal}{Phys. Rev. Lett.}}
  \textbf{\bibinfo{volume}{119}}, \bibinfo{pages}{123603}
  (\bibinfo{year}{2017}).

\bibitem{rossi_measurement-based_2018}
\bibinfo{author}{Rossi, M.}, \bibinfo{author}{Mason, D.},
  \bibinfo{author}{Chen, J.}, \bibinfo{author}{Tsaturyan, Y.} \&
  \bibinfo{author}{Schliesser, A.}
\newblock \bibinfo{title}{Measurement-based quantum control of mechanical
  motion}.
\newblock \eprint{Preprint at http://arxiv.org/abs/1805.05087 (2018).}

\bibitem{duan_long_2001}
\bibinfo{author}{Duan, L.-M.}, \bibinfo{author}{Lukin, M.~D.},
  \bibinfo{author}{Cirac, I.} \& \bibinfo{author}{Zoller, P.}
\newblock \bibinfo{title}{Long distance quantum communication with atomic
  ensembles and linear optics}.
\newblock \emph{\bibinfo{journal}{Nature}} \textbf{\bibinfo{volume}{414}},
  \bibinfo{pages}{413--418} (\bibinfo{year}{2001}).

\bibitem{rakhubovsky_squeezer_2016}
\bibinfo{author}{Rakhubovsky, A.~A.}, \bibinfo{author}{Vostrosablin, N.} \&
  \bibinfo{author}{Filip, R.}
\newblock \bibinfo{title}{{Squeezer-based pulsed optomechanical interface}}.
\newblock \emph{\bibinfo{journal}{Phys. Rev. A}} \textbf{\bibinfo{volume}{93}},
  \bibinfo{pages}{033813} (\bibinfo{year}{2016}).

\bibitem{zhang_quantum_2018}
\bibinfo{author}{Zhang, M.}, \bibinfo{author}{Zou, C.-L.} \&
  \bibinfo{author}{Jiang, L.}
\newblock \bibinfo{title}{Quantum transduction with adaptive control}.
\newblock \emph{\bibinfo{journal}{Phys. Rev. Lett.}}
  \textbf{\bibinfo{volume}{120}}, \bibinfo{pages}{020502}
  (\bibinfo{year}{2018}).

\bibitem{braunstein_teleportation_1998}
\bibinfo{author}{Braunstein, S.~L.} \& \bibinfo{author}{Kimble, H.~J.}
\newblock \bibinfo{title}{Teleportation of {Continuous} {Quantum} {Variables}}.
\newblock \emph{\bibinfo{journal}{Phys. Rev. Lett.}}
  \textbf{\bibinfo{volume}{80}}, \bibinfo{pages}{869--872}
  (\bibinfo{year}{1998}).

\bibitem{teufel_nanomechanical_2009}
\bibinfo{author}{Teufel, J.~D.}, \bibinfo{author}{Donner, T.},
  \bibinfo{author}{Castellanos-Beltran, M.~A.}, \bibinfo{author}{Harlow, J.~W.}
  \& \bibinfo{author}{Lehnert, K.~W.}
\newblock \bibinfo{title}{Nanomechanical motion measured with an imprecision
  below that at the standard quantum limit}.
\newblock \emph{\bibinfo{journal}{Nat. Nanotechnol.}}
  \textbf{\bibinfo{volume}{4}}, \bibinfo{pages}{820--823}
  (\bibinfo{year}{2009}).

\bibitem{peterson_laser_2016}
\bibinfo{author}{Peterson, R.~W.} \emph{et~al.}
\newblock \bibinfo{title}{Laser cooling of a micromechanical membrane to the
  quantum backaction limit}.
\newblock \emph{\bibinfo{journal}{Phys. Rev. Lett.}}
  \textbf{\bibinfo{volume}{116}}, \bibinfo{pages}{063601}
  (\bibinfo{year}{2016}).

\bibitem{andrews_2015}
\bibinfo{author}{Andrews, R.~W.}
\newblock \bibinfo{title}{\emph{Quantum signal processing with mechanical
  oscillators}}.
\newblock Ph.D. thesis (\bibinfo{year}{2015}).

\bibitem{menke_reconfigurable_2017}
\bibinfo{author}{Menke, T.} \emph{et~al.}
\newblock \bibinfo{title}{Reconfigurable re-entrant cavity for wireless
  coupling to an electro-optomechanical device}.
\newblock \emph{\bibinfo{journal}{Rev. Sci. Instrum.}}
  \textbf{\bibinfo{volume}{88}}, \bibinfo{pages}{094701}
  (\bibinfo{year}{2017}).

\bibitem{thompson_strong_2008}
\bibinfo{author}{Thompson, J.} \emph{et~al.}
\newblock \bibinfo{title}{Strong dispersive coupling of a high-finesse cavity
  to a micromechanical membrane}.
\newblock \emph{\bibinfo{journal}{Nature}} \textbf{\bibinfo{volume}{452}},
  \bibinfo{pages}{72--75} (\bibinfo{year}{2008}).
  
  \bibitem{peterson_laser_2016}
\bibinfo{author}{Peterson, R.~W.} \emph{et~al.}
\newblock \bibinfo{title}{Laser cooling of a micromechanical membrane to the
  quantum backaction limit}.
\newblock \emph{\bibinfo{journal}{Phys. Rev. Lett.}}
  \textbf{\bibinfo{volume}{116}}, \bibinfo{pages}{063601}
  (\bibinfo{year}{2016}).

\bibitem{andrews_bidirectional_2014}
\bibinfo{author}{Andrews, R.~W.} \emph{et~al.}
\newblock \bibinfo{title}{Bidirectional and efficient conversion between
  microwave and optical light}.
\newblock \emph{\bibinfo{journal}{Nat. Phys.}} \textbf{\bibinfo{volume}{10}},
  \bibinfo{pages}{321--326} (\bibinfo{year}{2014}).

\bibitem{purdy_cavity_2012}
\bibinfo{author}{Purdy, T.~P.}, \bibinfo{author}{Peterson, R.~W.},
  \bibinfo{author}{Yu, P.-L.} \& \bibinfo{author}{Regal, C.~A.}
\newblock \bibinfo{title}{Cavity optomechanics with {Si}$_3${N}$_4$ membranes
  at cryogenic temperatures}.
\newblock \emph{\bibinfo{journal}{New J. Phys.}} \textbf{\bibinfo{volume}{14}},
  \bibinfo{pages}{115021} (\bibinfo{year}{2012}).

\bibitem{aspelmeyer_cavity_2014}
\bibinfo{author}{Aspelmeyer, M.}, \bibinfo{author}{Kippenberg, T.~J.} \&
  \bibinfo{author}{Marquardt, F.}
\newblock \bibinfo{title}{Cavity optomechanics}.
\newblock \emph{\bibinfo{journal}{Rev. Mod. Phys.}}
  \textbf{\bibinfo{volume}{86}}, \bibinfo{pages}{1391--1452}
  (\bibinfo{year}{2014}).

\bibitem{hood_characterization_2001}
\bibinfo{author}{Hood, C.~J.}, \bibinfo{author}{Kimble, H.~J.} \&
  \bibinfo{author}{Ye, J.}
\newblock \bibinfo{title}{Characterization of high-finesse mirrors: {Loss},
  phase shifts, and mode structure in an optical cavity}.
\newblock \emph{\bibinfo{journal}{Phys. Rev. A}} \textbf{\bibinfo{volume}{64}},
  \bibinfo{pages}{033804} (\bibinfo{year}{2001}).

\bibitem{gao_noise_2007}
\bibinfo{author}{Gao, J.}, \bibinfo{author}{Zmuidzinas, J.},
  \bibinfo{author}{Mazin, B.~A.}, \bibinfo{author}{LeDuc, H.~G.} \&
  \bibinfo{author}{Day, P.~K.}
\newblock \bibinfo{title}{Noise properties of superconducting coplanar
  waveguide microwave resonators}.
\newblock \emph{\bibinfo{journal}{Appl. Phys. Lett.}}
  \textbf{\bibinfo{volume}{90}}, \bibinfo{pages}{102507}
  (\bibinfo{year}{2007}).

\bibitem{gao_experimental_2008}
\bibinfo{author}{Gao, J.} \emph{et~al.}
\newblock \bibinfo{title}{Experimental evidence for a surface distribution of
  two-level systems in superconducting lithographed microwave resonators}.
\newblock \emph{\bibinfo{journal}{Appl. Phys. Lett.}}
  \textbf{\bibinfo{volume}{92}}, \bibinfo{pages}{152505}
  (\bibinfo{year}{2008}).

\bibitem{zhang_quantum_2018}
\bibinfo{author}{Zhang, M.}, \bibinfo{author}{Zou, C.-L.} \&
  \bibinfo{author}{Jiang, L.}
\newblock \bibinfo{title}{Quantum transduction with adaptive control}.
\newblock \emph{\bibinfo{journal}{Phys. Rev. Lett.}}
  \textbf{\bibinfo{volume}{120}}, \bibinfo{pages}{020502}
  (\bibinfo{year}{2018}).

\bibitem{gottesman_encoding_2001}
\bibinfo{author}{Gottesman, D.}, \bibinfo{author}{Kitaev, A.} \&
  \bibinfo{author}{Preskill, J.}
\newblock \bibinfo{title}{{Encoding a qubit in an oscillator}}.
\newblock \emph{\bibinfo{journal}{Phys. Rev. A}} \textbf{\bibinfo{volume}{64}},
  \bibinfo{pages}{012310} (\bibinfo{year}{2001}).

\bibitem{albert_performance_2017}
\bibinfo{author}{Albert, V.~V.} \emph{et~al.}
\newblock \bibinfo{title}{Performance and structure of single-mode bosonic
  codes}.
\newblock \emph{\bibinfo{journal}{Phys. Rev. A}} \textbf{\bibinfo{volume}{97}},
  \bibinfo{pages}{032346} (\bibinfo{year}{2018}).

\end{thebibliography}
\end{document}